\documentclass[a4paper,11pt]{article}
\pdfoutput=1

\usepackage{jheppub}
\usepackage{amsmath,amssymb}
\usepackage{booktabs}
\usepackage[utf8]{inputenc}
\usepackage{hhline}
\usepackage{diagbox}
\usepackage{multirow}
\usepackage{subfigure}
\usepackage{float}
\usepackage{tabularx}
\usepackage{cleveref}
\usepackage[dvipsnames]{xcolor}
\usepackage[normalem]{ulem}
\usepackage{cancel}
\newcommand{\ba}{\begin{eqnarray}}
\newcommand{\ea}{\end{eqnarray}}
\newcommand{\be}{\begin{equation}}
\newcommand{\ee}{\end{equation}}
\newcommand{\bal}{\begin{align}}
\newcommand{\eal}{\end{align}}

\newcommand{\eT}{\eta}
\newcommand{\eP}{\eta^{\prime}}

\makeatletter
\def\fmslash{\@ifnextchar[{\fmsl@sh}{\fmsl@sh[0mu]}}
\def\fmsl@sh[#1]#2{%
  \mathchoice
    {\@fmsl@sh\displaystyle{#1}{#2}}%
    {\@fmsl@sh\textstyle{#1}{#2}}%
    {\@fmsl@sh\scriptstyle{#1}{#2}}%
    {\@fmsl@sh\scriptscriptstyle{#1}{#2}}}
\def\@fmsl@sh#1#2#3{\m@th\ooalign{$\hfil#1\mkern#2/\hfil$\crcr$#1#3$}}
\makeatother

\definecolor{darkgreen}{RGB}{90,150,50}
\definecolor{brown}{RGB}{150,50,0}

\definecolor{blue}{rgb}{0,0,0.5}
\definecolor{darkgreen}{RGB}{0,175,10}
\definecolor{brown}{RGB}{150,50,0}

\title{Form factors and phenomenology of $\boldsymbol{B_{(s)}}$ and $\boldsymbol{D_{(s)}}$ semileptonic decays to $\boldsymbol{\eta}$ and $\boldsymbol{\eta^\prime}$}

\author[a]{Bla\v zenka  Meli\'c,}
\author[b]{M\'eril Reboud}

\affiliation[a]{Rudjer Boskovic Institute, Division of Theoretical Physics, Bijeni\v cka 54, HR-10000 Zagreb, Croatia}
\affiliation[b]{Université Paris-Saclay, CNRS/IN2P3, IJCLab, 91405 Orsay, France}

\emailAdd{%
melic@irb.hr,
meril.reboud@cnrs.fr%
}

\abstract{
Motivated by more precise recent measurements of the $B \to \eta \ell \nu$ and $D_{(s)} \to \eta^{(\prime)}\ell \nu$ decays, we employ state-of-the-art parametrizations to describe the $B_{(s)}, D_{(s)} \to \eta^{(\prime)}$ form factors across the full $q^2$ range, fitting them to the latest light-cone sum rule results. 
Using these results, we compute the branching ratios for all relevant decays and compare them with experimental data, finding good agreement.
Additionally, we examine the validity and precision of the extracted \(\eta\)-\(\eta'\) mixing angle in the context of heavy meson decays.
By combining our predictions for the $B, D, D_s \to \eta^{(\prime)}$ form factors with recent semileptonic decay measurements, we extract the CKM elements $|V_{ub}|$, $V_{cs}$, and $V_{cd}$.
We also provide tests of the lepton flavour universality in the $B, D, D_s \to \eta^{(\prime)}\ell \nu$ decays and present results for the forward-backwards asymmetries in these decays.
Our findings indicate that the extracted values are becoming comparable in precision to those obtained from more conventional semileptonic decay analyses.
}

\begin{document}
\preprint{RBI-ThPhys-2025-29, EOS-2025-04}
\maketitle
\flushbottom


\section{Introduction}
The decays of heavy mesons to $\eT$ and $\eP$ provide valuable insights into $\eT-\eP$ phenomenology and mixing, making them important systems for studying these effects.
Moreover, the precise knowledge of semileptonic branching ratios (BRs) of $B_{(s)}, D_{(s)} \to \eta^{(\prime)} \ell \nu$ decays plays a major role in the extraction of CKM matrix elements. 
Notably, the BRs of $B \to \eta^{(\prime)} \ell \nu$ are used for background estimation in the extraction of $V_{ub}$ from $B \to \pi \ell \nu$ decays, while $D_{(s)} \to \eta^{(\prime)} \ell \nu$ decays have already achieved sufficient precision to allow the extraction of $V_{cs}$ and $V_{cd}$ matrix elements.
These modes are also used in the measurement of inclusive semileptonic $B \to X_u \ell \nu$ decays, where the hadronic final state is modelled as a combination of exclusive, resonant semileptonic modes $ B \to \{ \pi, \eta, \eta^{\prime}, \omega, \rho\} \ell \nu$, and non-resonant contributions (see \cite{Belle:2021ymg}). The precision of these modes is important not only for determining $|V_{ub}|_{\rm inc}$, but also because the differential distributions provide valuable input for studying non-perturbative shape functions \cite{Neubert:1993ch}.
In addition, the form factors of $B_{(s)}, D_{(s)} \to \eta^{(\prime)}$ transitions are essential for $D_{(s)},B_{(s)} \to P P$ and $B_{(s)}\to X_{c\bar{c}} P$ ($P = \pi, K, \eta, \eta'$) phenomenology, i.e. for understanding the factorization hypothesis and the origin of the nonfactorizable contributions in these decays \cite{Beneke:2003zv,Colangelo:2010wg,Fleischer:2011ib, LHCb:2014oms}. \\

The main ingredients of the paper are as follows. 

Firstly, we improve the form factor analysis of semileptonic $B_{(s)}$ and $D_{(s)}$ decays to $\eT$ and $\eP$ to bring these form factors to the same level of scrutiny as it was done for semileptonic $B \to \pi$ \cite{Leljak:2021vte} and $B_{s} \to K$ and $D$ \cite{Bolognani:2023mcf} transitions.
In our work, the  $B_{(s)}, D_{(s)} \to \eta^{(\prime)}$ form factors are calculated using LCSR by improving the work done in Ref.~\cite{Duplancic:2015zna}, using new values of heavy quark/meson parameters and new lattice QCD results on  $\eT$ and $\eP$ parameters and their distribution amplitudes from \cite{Bali:2021qem} (see also \cite{CSSMQCDSFUKQCD:2021rvs} and more recently from the same UKQCD coll \cite{CSSMQCDSFUKQCD:2023rei}).
Apart from these changes, the main improvements are in the $q^2$ parametrisation of the form factors, which are performed using a simplified series expansion \cite{Boyd:1994tt,Bharucha:2010im}.
The statistical analysis is done using the EOS software v1.0.17~\cite{EOSAuthors:2021xpv,EOS:v1.0.17}, which allows for a detailed Bayesian analysis of the form factors and their phenomenology.
We then compare our results with the first lattice QCD calculation of the unphysical $B_s \to \eta_s$ decay \cite{Parrott:2020vbe}.

Secondly, we perform a phenomenological analysis of the $B \to \eta^{(\prime)} \ell \nu$ and $D_{(s)} \to \eta^{(\prime)} \ell \nu$ semileptonic decays using the improved form factors.
We provide the BRs for these decays, which are then examined relative to the latest experimental results.
By combining our form factors with the most recent measurements of $B \to \eT$ and $D_{(s)} \to \eta^{(\prime)}$, we extract the $|V_{ub}|, V_{cs}$ and $V_{cd}$ CKM matrix elements and compare them with those extracted from other semileptonic decays. 
In addition, BESIII has reported measurements of some of the forward-backwards asymmetries, $\mathcal{A}_\mathrm{FB}$, and lepton-flavour universality (LFU) ratios, which we examine from a theoretical perspective.

Finally, with the precisely extracted form factors and the corresponding analytical expressions, we are well-positioned to validate the extraction of the $\eT-\eP$ mixing angle from data, as commonly done in the literature.

\section{Properties of $\eT$ and $\eP$ in the LCSR calculation of the $B_{(s)}, D_{(s)} \to \eT,\eP$ form factors}

Here, we identify the main characteristics of the heavy meson semileptonic decays to $\eT$ and $\eP$ particles important for our analysis.

Understanding $\eta$ and $\eta^{\prime}$ mesons implies an understanding of the nonperturbative QCD dynamics and local symmetries due to their quark and gluon content and their mixing. 
The properties and the rich phenomenology of $\eT$  and $\eP$ were extensively discussed in the past (see \cite{Bass:2018xmz} and references therein). 

In the SU(3) flavour-symmetry limit, the $\eT$ particle corresponds to the flavour-octet member of the Goldstone particles, $\eta_8  = 1/\sqrt{6} ( u\bar{u} + d \bar{d} - 2 s \bar{s})$ and $\eP$ is $\eT_0  = 1/\sqrt{6} ( u\bar{u} + d \bar{d}+ s \bar{s})$, a singlet with a much larger mass which is attributed to the gluonic mass term stemming from the non-perturbative dynamics via the $U_A(1)$ anomaly.
The anomalous gluonic term, represented by the topological charge density, violates the conservation of the flavour-singlet axial current, even in the chiral limit.
In the SU(3) limit,  without the gluonic anomaly, the physical $\eT$ particle would be an isosinglet light-quark state, $|\eT \rangle  = 1/\sqrt{2}|u\bar{u} + d \bar{d} \rangle$ with a mass $m_{\eT}^2 \sim m_{\pi}^2$ and the physical $\eP$ would be just a strange-quark state, $|\eP \rangle = |s\bar{s} \rangle$ having a mass $m_{\eP}^2 \sim (2 m_K^2 - m_{\pi}^2)$, according to the leading order chiral expansion.

Since the SU(3) corrections in $\eT$ and $\eP$ physics are empirically confirmed to be large, it is difficult to describe correctly the physical states of these particles.
Going beyond the leading order chiral expansion introduces SU(3) breaking through the difference in the pion and kaon constants, and some additional OZI-violating corrections appear.

In full generality, the mixing of the physical $\eT$ and $\eP$ particles can be described in the singlet-octet (SO) base or in the quark-flavour (QF) base.
Having in mind the nature of the particles, the SO-base looks more natural since it naturally separates the $U(1)_A$ axial anomaly in the singlet decay constant $f_1$, which renormalises multiplicatively, while the octet decay constant $f_8$ is scale-independent, as well as the angles $(\theta_1, \theta_8)$.
Therefore, we have
\ba
f_P^8 (\mu) &=& f_P^8 (\mu_0)\,, \nonumber \\
f_P^1 (\mu) &=& f_P^1 (\mu_0) \left ( 1 + \frac{2 n_f}{\pi \beta_0} \left [ \alpha_s(\mu) - \alpha_s(\mu_0) \right ] \right )\,,
\ea
where $P = \eT, \eP$, $\mu_0 = 1$ GeV is the scale at which the values of the mixing parameters are determined \cite{Feldmann:1998vh}, and 
\be
\left ( \begin{matrix}
 f_\eta^8 & f_\eta^1 \\
 f_{\eta^\prime}^8  & f_{\eta^\prime}^1 \\
 \end{matrix}
 \right )  = \left ( \begin{matrix}
 \cos \theta_8 & - \sin \theta_1 \\
 \sin \theta_8  &  \cos \theta_1 \\
 \end{matrix}
 \right ) \left ( \begin{matrix}
 f_8 & 0 \\
 0  &  f_1 \\
 \end{matrix}  \right ) \,.
\ee
The phenomenological fit of these parameters was performed in \cite{Feldmann:1998vh,Feldmann:1999uf} with the values: 
\ba
f_8 = (1.26 \pm 0.04) f_{\pi}, \qquad \theta_8 = -21.2^{\circ} \pm 1.6^{\circ} \nonumber \\
f_1 = (1.17 \pm 0.03) f_{\pi}, \qquad \theta_1 = -9.2^{\circ} \pm 1.7^{\circ} \,.
\ea
The newest lattice results \cite{Bali:2018spj, Ottnad:2017bjt, Ottnad:2025zxq} confirm the range of the parameters. 
On the other hand, the QF base looks more intuitive since it is defined as
\ba
\left ( \begin{matrix}
 f_\eta^q & f_\eta^s \\
 f_{\eta^\prime}^q  & f_{\eta^\prime}^s \\
 \end{matrix}
 \right )  = \left ( \begin{matrix}
 \cos \phi_q & - \sin \phi_s \\
 \sin \phi_q  &  \cos \phi_s \\
 \end{matrix}
 \right ) \left ( \begin{matrix}
 f_q & 0 \\
 0  &  f_s \\
 \end{matrix}  \right ) \,,
 \label{eq:fetas}
\ea
in terms of the decay constants of pure (hypothetical) non-strange and strange states, $f_q$ and $f_s$, respectively. 
In this basis, the angles are scale-dependent, but their difference is defined by the OZI-rule violating contributions, which are known to be small.
Therefore, the approximation 
\be
\phi_q = \phi_s = \phi \,,
\ee
was introduced in Ref.~\cite{Feldmann:1998vh} and significantly simplifies the picture and the treatment of the $\eT$-$\eP$ mixing just in terms of three parameters.
This scheme is usually referred to as the Feldmann-Kroll-Stech (FKS) scheme.
The values extracted in this scheme are~\cite{Feldmann:1998vh}
\be
f_q = (1.07 \pm 0.02 ) f_{\pi} \,, \qquad
f_s = (1.34 \pm 0.06 ) f_{\pi} \,, \qquad
\phi = 39.3^{\circ} \pm 1.0^{\circ} \,.
\label{eq:FKS}
\ee
The usage of a single mixing angle is supported by the most recent lattice study~\cite{Bennett:2024wda}.
The previous detailed study of the FKS parameters on the lattice \cite{Bali:2021qem} confirmed the values cited above, within uncertainties, along with the latest lattice study in Ref.~\cite{Ottnad:2025zxq}.
Therefore, in this paper, we will use values from \cref{eq:FKS} as our default values for these parameters.

Gluonic contributions can also impact $\eta$ and $\eP$ physics in several other ways, such as mixing with the lightest glueball state $|G \rangle$ and playing a role in processes involving high-momentum transfer. 
The most recent lattice study of the mixing of the flavour singlet pseudoscalar meson $\eta$ and the pseudoscalar glueball $G$ in the $N_f = 2$ QCD \cite{Jiang:2022ffl} suggests that the mixing effects can be safely neglected when the properties of $\eta$ and $\eta^{\prime}$ are considered and that the origin of $\eta$ mass is predominantly coming from the $U(1)_A$ anomaly\footnote{The similar conclusion was derived from phenenomenological consideration in \cite{Zhang:2025yeu}.}.
However, in the latter case, perturbative QCD allows for the possibility of radiatively generated gluons contributing to the formation of the $\eT$ and $\eP$  mesons through a two-gluon distribution amplitude.
This effect has to be included in our calculation of the form factors, as it will be shown below. 
In addition, the mixing with other pseudoscalar mesons, particularly the pion, can be taken into account.
This mixing was shown to be negligible, although a visible $\pi$-content is observed in $\eP$~\cite{CSSMQCDSFUKQCD:2021rvs}.

Following these considerations on the complexity of the $\eT$-$\eP$ physics, the approximations used in the calculation of the $B_{(s)}, D_{(s)} \to \eT$, $\eP$ form factors have to be carefully examined to effectively capture the full range of effects, but with reasonable approximations, which will reduce the uncertainty due to the lack of precise knowledge of all phenomenological $\eT$ and $\eP$ parameters.\\

We extract the $B_{(s)}, D_{(s)} \to \eta, \eta^{\prime}$ transition form factors in the LCSR framework, in the setup where the $\eT$ and $\eP$ particles are described by the light-cone distribution amplitudes (DAs) of various twists.

The leading twist DAs of the $\eT$ and $\eP$ mesons $(P = \eT, \eP)$ are defined as the twist-2 two-quark DAs of $\eta^{(\prime)}$.
Being symmetric in their argument, they can be expanded in terms of Gegenbauer polynomials
\ba
\psi_{2,P}^i(u) = 6 u (1-u) \left ( 1 + \sum_{n=2,4,..} a_n^{P,i}(\mu) C_n^{3/2}(2u -1) \right )\, \quad (i = 1,8,q,s)\,.
\ea 
The coefficients $a_n^{P,i}$ are the $n$th Gegenbauer moments of the quark DA. 

At that twist-level, there is also the gluonic twist-2 $\psi_{2,P}^g(u)$ DA contributing, which is defined by the following matrix element \cite{Baier:1981pm,Shifman:1980dk}:
\ba
n_{\mu} n_{\nu} \langle 0| G^{\mu\alpha}(z) [z,-z] \tilde{G}_{\alpha}^{\nu}(-z) | P(p) \rangle = 
\frac{1}{2} \frac{C_F}{\sqrt{3}} (pz)^2 f_P^i \int_0^1 du e^{i (2u -1) (pz)} \psi_{2,P}^g(u)\,.
\ea
It is antisymmetric and therefore 
\ba
\psi_{2,P}^g(u) =  - \psi_{2,P}^g(1-u) \,,
\ea
and
\ba
\psi_{2,P}^g(u) = u^2 (1-u)^2 \left ( \sum_{n=2,4,..} b_n^{P,g}(\mu) C_{n-1}^{5/2}(2u -1) \right ) \,,
\ea
where the coefficients $b_n^{P,g}$ are unknown Gegenbauer moments of the gluon DA, and we take $b_n^{\eta,g} = b_n^{\eta^{\prime},g} = b_n^{g}$. For the same reason, we will also take just $b_2^{g}$ into consideration. 

Generally, we have in the FKS scheme
\ba
|\eta_q\rangle  \propto \psi_2^q (u) | q\bar{q} \rangle + \psi_2^\mathrm{OZI} |s \bar{s} \rangle + \psi_2^g (u) | G\bar{G} \rangle + ...
\nonumber \\
|\eta_s\rangle  \propto \psi_2^s (u) | s\bar{s} \rangle + \psi_2^\mathrm{OZI} |q \bar{q} \rangle + \psi_2^g (u) | G\bar{G} \rangle + ...
\ea
The simplified FKS picture of $\eT-\eP$ mixing also completely removes the scale dependence of the parameters. Namely, for consistency with neglecting OZI-violating contributions in the mixing, one should also consistently use, in the twist-2 part,  
\be
\psi_2^\mathrm{OZI} = \frac{\sqrt{2}}{3} (\psi_2^1 - \psi_2^8) = 0 \,,
\ee
i.e 
$\psi_2^8 = \psi_2^1$ for the DAs. 
Due to the simplification in the QF-base, we also have%
\footnote{Usually the equality is taken at the hadronic scale $\mu = 1$ GeV and with the running of $a_2$ by the scaling-low of the octet Gegenbauer coefficient, 
$a_2(\mu)= \left ( \alpha_s(\mu)/\alpha_s(\mu_0) \right )^{50/(9 \beta_0)} a_2^8(1 \, {\rm GeV}) $.  }
\be
a_2^1(\mu) = a_2^8(\mu)
\ee
and consequently 
\be
a_2^q = a_2^s\,.
\ee
Looking at the lattice results on the parameters, $a_2^{\pi} = 0.116^{+0.016}_{-0.020}$, $a_2^K = 0.106^{+0.015}_{-0.016}$ \cite{RQCD:2019osh}, there is no evidence for a large SU(3) breaking and the approximation is meaningful. In the following we assume numerically $a_2^q = a_2^{\pi}$.

However, we know that this cannot be fully true since these DAs have different scale dependence and evolve differently.
Moreover, we also know that there is a gluonic contribution to the $\eT$ and $\eP$ states. 
With the above normalisation of the DA, the renormalisation mixing of twist-2 quark and gluonic distribution amplitudes is given as \cite{Kroll:2002nt,Ball:2007hb,Agaev:2014wna}
\ba
\mu \frac{d}{d \mu} 
\left ( \begin{matrix} a_2^{\eta^{(\prime)},1} \\ b_2^{\eta^{(\prime)} ,g}\\ \end{matrix} \right ) 
= -\frac{\alpha_s(\mu)}{4 \pi} \left ( \begin{matrix}
 \frac{100}{9} & - \frac{10}{81}  \\
 -36   &  22 \\
 \end{matrix}
 \right ) \left ( \begin{matrix} a_2^{\eta^{(\prime)},1} \\ b_2^{\eta^{(\prime)} ,g} \\ \end{matrix} \right ) \,,
\label{eq:evol}
\ea
and it is numerically small.
The mixing is however important, since it verifies the collinear ``factorisation formula'' for the form factors
\ba
F(q^2, (p+q)^2)  = \int_0^1 du\,\sum_n T_H^{(n)} (u, q^2, (p+q)^2,\mu_{\rm IR}) \psi_{n}(u,\mu_{\rm IR})\,, 
\ea
and proves that the separation of the transition form factors in perturbatively calculable hard-scattering $T_H$ part and a nonperturbative DA is essentially independent of the factorization scale $\mu_{\rm IR}$ \cite{Melic:2001wt}. 
If we fully neglect the evolution of the singlet amplitude, the factorization in the $B, D \to \eT,\eP\ell \nu$ processes cannot be proven at the NLO, since the $\mu$-evolution of $\phi_2 T_H$ exactly cancels with $\phi_2^g T_H^g$, when scaled properly \cite{Duplancic:2015zna, Ball:2007hb}. 

Therefore, we have to consider mixing among the twist-2 quark and gluon DAs, as well as the complete running of $a_2$, although the effect is numerically small%
\footnote{Ref.~\cite{Porkert:2019okj} analysed the impact of the RG running of the Gegenbauer moments and found it to be under control within the physically relevant momentum range of $[1.0, 2.5]$ GeV.}.
On the other hand, due to the lack of better knowledge on the gluon content in $\eta$ and $\eP$, we are forced to treat $b_2^{\eta,g}= b_2^{\eta',g} \equiv b_2^g$ as a free parameter, usually ranging from $-20$ to $+20$ \cite{Ball:2007hb}.

We truncated the Gegenbauer expansion of twist-2 quark and gluon DAs to the first couple of coefficients, $a_2$, $a_4$ and $b_2$.\\

The $\eT$ and $\eP$ DAs of higher twist are defined following \cite{Agaev:2014wna} and \cite{Ball:2006wn}.
Their parameter evolutions and definitions include now the anomaly contribution $a_P$ with the following expressions in the FKS scheme \cite{Beneke:2002jn}: 
\ba
a_\eta = - \frac{1}{\sqrt{2}} ( f_q m_\eta^2 - h_q) \cos\phi = - \frac{ m_{\eta^{\prime}}^2 -m_\eta^2}{\sqrt{2}} \sin\phi \cos\phi \left ( - f_q \sin\phi + \sqrt{2} f_s \cos\phi \right) \,, \nonumber \\
a_{\eta^{\prime}} = - \frac{1}{\sqrt{2}} ( f_q m_{\eta^{\prime}}^2 - h_q) \sin\phi = - \frac{ m_{\eta^{\prime}}^2 -m_\eta^2}{\sqrt{2}} \sin\phi \cos\phi \left (f_q \cos\phi + \sqrt{2} f_s \sin\phi \right)\,. \nonumber \\
\ea
If the anomalies are neglected, the following approximations in the twist-3 and twist-4 DAs are valid: 
\be
 f_{\pi} \frac{m_{\pi}^2}{2 m_q} \to f_{q} \frac{m_{\pi}^2}{2 m_q} \,, \qquad 
 f_{\pi} \frac{m_{\pi}^2}{2 m_s} \to f_{s} \frac{2m_{K}^2 - m_{\pi}^2}{2 m_s}  \,
\label{eq:hq-approx} 
\ee
for $H \to \eta_q$ and $H \to \eta_s$ decays, respectively. Instead, we are going to use (in the FKS scheme): 
\begin{align}\label{eq:HqHs}
& f_{\pi} m_{\pi}^2 \to h_{q} = f_q (m_\eta^2 \cos^2 \phi + m_{\eta^{\prime}}^2 \sin^2 \phi) - \sqrt{2} f_s (m_{\eta^{\prime}}^2 - m_\eta^2) \sin \phi \cos \phi \,, \nonumber \\
& f_{\pi} m_{\pi}^2 \to h_{s} = f_s (m_{\eta^{\prime}}^2 \cos^2 \phi + m_{\eta}^2 \sin^2 \phi) - \frac{f_q}{\sqrt{2}} (m_{\eta^{\prime}}^2 - m_\eta^2) \sin \phi \cos \phi  \, .  
\end{align} 
The above quantities, especially $h_q$, are weakly constrained due to the numerical cancellations.
Varying $\phi, f_q$ and $f_s$ in the ranges given in \cref{eq:FKS}, we obtain
\be
h_{q} = 0.0017 \pm 0.0037 \,, \quad h_{s} = 0.0871 \pm 0.0056.
\label{eq:hqhs} 
\ee

The mixing angle $\phi$-dependence of $h_q$ and $h_s$ in \cref{eq:HqHs} will, as the leading twist-2 mixing of the quark and gluonic DAs, break the simple direct proportionality of the form factors to $\cos\phi$ and $\sin\phi$.
This is due to the $U(1)_A$ anomaly corrections in the twist-3 and twist-4 DAs. However, since the higher-twist corrections are numerically small, this effect in the form factors is found to be negligible in the physical $\phi$ region.
Outside of this region, $h_q$ becomes unphysically large, and the entire twist expansion breaks down. 

The $\eT$, $\eP$ decay constants and other relevant parameters summarised in the \cref{app:fixed_params} are taken from \cite{Bali:2021qem}, agreeing with the values provided by other lattice collaborations \cite{Ottnad:2025zxq,CSSMQCDSFUKQCD:2023rei}.

The main physics discussion and all relevant LCSR expressions can be found in \cite{Duplancic:2015zna} and \cite{Ball:2007hb,Offen:2013nma}, including the NLO LCSR expressions of the form factors from \cite{Duplancic:2008ix, Duplancic:2008tk}. 

\section{Analysis of  $B_{(s)} \to \eT, \eP$ and $D_{(s)} \to \eT, \eP$ form factors}

The hadronic matrix elements defining the transitions $B \to \eT, \eP$ can be written as
\be
\langle \eta^{(\prime)}(p)|\bar{u} \gamma_\mu b |\bar B(p+q)\rangle=
2f^+_{B\eta^{(\prime)}}(q^2)p_\mu +\left(f^+_{B\eta^{(\prime)}}(q^2)+f^-_{B\eta^{(\prime)}}(q^2)\right)q_\mu\,,
\label{eq:fplBpi}
\ee
\be
\langle \eta^{(\prime)}(p)|\bar{u} \sigma_{\mu \nu}q^\nu b
|\bar B(p+q)\rangle=
\Big [q^2(2p_\mu+q_\mu) - (m_B^2-m_{\eta^{(\prime)}}^2) q_\mu\Big ]
\frac{i f_{B\eta^{(\prime)}}^T(q^2)}{m_B+m_{\eta^{(\prime)}}}\,.
\label{eq:fsigBpi}
\ee
and similarly for the $B_s, D$, and $D_s$ transitions%
\footnote{%
In the literature it sometimes appears that the form factors are defined as above by divided by a factor $\sqrt{2}$ to match the transition form factors of $\eta,\eta^\prime$ with those of a pion when there is no $\eta-\eta^\prime$ mixing and in the limit of the conserved SU(3)-flavour symmetry \cite{Ball:2007hb,Gonzalez-Solis:2018ooo}. 
}
to $\eT$ and $\eP$. 
The scalar $B \to \eta^{(\prime)}$ form factor is then a combination of the vector form factors,
\be
f^0_{B\eta^{(\prime)}}(q^2) = f^+_{B\eta^{(\prime)}}(q^2) + \frac{q^2}{m_B^2-m_{\eta^{(\prime)}}^2} f^-_{B\eta^{(\prime)}}(q^2) 
\label{eq:f0}
\ee
and it only appears in the semileptonic $B_{(s)}, D_{(s)} \to \eta^{(\prime)} \ell \nu$ decays when the lepton mass is not neglected, and in the rare $B_{s}, D_{s} \to \eta^{(\prime)} \ell^+ \ell^-$ and $\nu \bar{\nu}$  decays. 

The form factors are calculated by the standard LCSR method, and our evaluation closely follows the work from \cite{Duplancic:2015zna}. 
The final expressions in the LCSR have the form 
\ba
\hspace*{-2cm} && f^+_{B\eta^{(\prime)}}(q^2) = \frac{e^{m_B^2/M^2}}{2m_B^2 f_B}
\Bigg[F_{0,B\eta^{(\prime)}}(q^2,M^2,s_0^B)+
\frac{\alpha_s C_F}{4\pi}\left ( F_{1,B\eta^{(\prime)}}(q^2,M^2,s_0^B) +  F_{1,B\eta^{(\prime)}}^{gg,+}(q^2,M^2,s_0^B) \right )
\Bigg]\,,
\nonumber \\
\label{eq:fplusLCSR}
\\
&& f^+_{B\eta^{(\prime)}}(q^2)+f^-_{B\eta^{(\prime)}}(q^2) = 
\frac{e^{m_B^2/M^2}}{m_B^2 f_B}
\Bigg[\widetilde{F}_{0,B\eta^{(\prime)}}(q^2,M^2,s_0^B)+
\frac{\alpha_s C_F}{4\pi}\widetilde{F}_{1,B\eta^{(\prime)}}(q^2,M^2,s_0^B)
\Bigg]\,,
\label{eq:fplminLCSR}
\\
\hspace*{-1cm}&& f^T_{B\eta^{(\prime)}}(q^2) =
\frac{(m_B+m_\eta^{(\prime)})e^{m_B^2/M^2}}{2m_B^2 f_B}
\Bigg[F^T_{0,B\eta^{(\prime)}}(q^2,M^2,s_0^B)
\nonumber \\
&& \hspace*{3cm} + \frac{\alpha_s C_F}{4\pi} \left ( F^T_{1,B\eta^{(\prime)}}(q^2,M^2,s_0^B) +  F_{1,B\eta^{(\prime)}}^{gg,T}(q^2,M^2,s_0^B) \right )
\Bigg]\,.
\label{eq:fTLCSR}
\ea
$f_B$ is the decay constant of the $B$-meson, and $s_0$ and $M^2$ are the effective threshold parameter and the Borel parameter, respectively, and will be determined later, for each of the form factors separately.

The form factors can be represented as
\ba
\left ( \begin{matrix}
 f_{B\eta}^{+,0,T}\\
  f_{B\eta^{\prime}}^{+,0,T} \\
 \end{matrix}
 \right )  = U(\phi) \left ( \begin{matrix}
 f_{B\eta^{q}}^{+,0,T}  \\
f_{B\eta^{s}}^{+,0,T} \\
 \end{matrix}  \right ) \,,
\ea
where
\ba
f_{B\eta^{q}}^{+,0,T}  = f_{B\eta^{q}}^{(\bar{q}q)\,+,0,T} + f_{B\eta^{q}}^{(gg)\,+,0,T}, \qquad 
f_{B\eta^{s}}^{+,0,T}  = 
f_{B\eta^{s}}^{(\bar{s}s)\,+,0,T} +f_{B\eta^{s}}^{(gg)\,+,0,T} \,.
\ea
In our setup, these objects still have a non-trivial $\phi$ dependence%
\footnote{The non-trivial $\phi$ dependence and additional sources of SU(3)-breaking were recently discussed and parametrically studied in Ref.~\cite{Bolognani:2024zno} in the context of non-leptonic $D$ decays.}%
,stemming from $U(1)_A$ higher-twist corrections (\cref{eq:HqHs}) and from $O(\alpha_s)$ gluonic term proportional to $f_\eta^1$ and $f_{\eta^\prime}^1$.

Explicitly, we have
\ba
f_{B\eta}^{+,0,T} = \frac{f_\eta^{(q)}}{\sqrt{2}} \left ( F_0^{(\bar{q}q)} +  F_1^{(\bar{q}q)}\right )^{+,0,T} + f_\eta^1 F_1^{(gg)\,+,0,T} \,,
\nonumber \\
f_{B\eta^\prime}^{+,0,T} = \frac{f_{\eta^\prime}^{(q)}}{\sqrt{2}} \left ( F_0^{(\bar{q}q)} +  F_1^{(\bar{q}q)}  \right ) ^{+,0,T}+ f_{\eta^\prime}^1 F_1^{(gg)\,+,0,T} \,,
\nonumber \\
f_{B_s\eta}^{+,0,T} =f_\eta^{(s)} \left ( F_0^{(\bar{s}s)} +  F_1^{(\bar{s}s)}  \right )^{+,0,T} + f_\eta^1 F_1^{(gg)\,+,0,T} \,,
\nonumber \\
f_{B_s\eta^\prime}^{+,0,T} = f_{\eta^\prime}^{(s)}\left ( F_0^{(\bar{s}s)} +  F_1^{(\bar{s}s)}  \right )^{+,0,T} + f_{\eta^\prime}^1 F_1^{(gg)\,+,0,T} \,,
\label{eq:fftotal}
\ea
where $F_0^{(\bar{q}q)}$ and $F_0^{(\bar{s}s)}$ ($F_1^{(\bar{q}q)}$ and $F_1^{(\bar{s}s)}$) are LO (NLO) contributions from quark hard-scattering amplitudes%
\footnote{For the compactness of the notation, for the scalar form factors we have changed the notation $\tilde F_{0,1}^{(\bar{q}q)},\tilde F_{0,1}^{(\bar{s}s)}$ by $ F_{0,1}^{(\bar{q}q)\,0},F_{0,1}^{(\bar{s}s)\,0}$ in  (\ref{eq:fftotal}) above.}%
for each of the form factors and $F_1^{(gg)}$ is the NLO gluonic contribution proportional to the singlet-flavour decay constants
\ba
f_\eta^1 &=& \frac{1}{\sqrt{3}} \left ( \sqrt{2} \cos\phi f_q - \sin\phi f_s \right )  \,,
\nonumber \\
f_{\eta^\prime}^1 &=& \frac{1}{\sqrt{3}} \left ( \sqrt{2} \sin\phi f_q + \cos\phi f_s \right ) \,, 
\ea
and the $f_{\eta^(\prime)}^{(r)}$ decay constants are given in (\ref{eq:fetas}). 
Analogous expressions are valid for $D_{(s)} \to \eta^{(\prime)}$ decays.
 
For $B, D \to \eta^{(\prime)}$ transitions the main contribution comes from $\eta_q$ meson states and $\eta_s$ contributes only through suppressed gluonic contributions, while for $B_s, D_s \to \eta^{(\prime)}$ transitions the leading $\eta_s$ meson state contribution will receive, through the gluonic diagrams, a small mixture with $\eta_q$ state.  Also, implicitly there will be mixing among twist-2 quark and gluonic distribution amplitudes, Eq.(\ref{eq:evol}), which will bring $b_2^{\eta^{(\prime),g}}$ dependence in the twist-2 quark LO ($F_0^{\bar{q}q}$ and $F_0^{\bar{s}s}$ ) and NLO contributions  ($F_1^{\bar{q}q}$ and $F_1^{\bar{s}s}$) and $a_2^{\eta^{(\prime),1}}$ dependence to the gluonic contributions $F_1^{gg}$. Moreover, higher twist contributions will bring $U(1)_A$ anomaly effects by higher-twist DAs' dependence on $h_q$ and $h_s$ parameters. 
The gluonic contributions, which are already NLO effects, will be calculated for $p^2 = m_{\eta^{(\prime)}}^2 = 0$.  
 
The form factors are calculated in the $\overline{MS}$ scheme, by taking the DAs of $\eT$ and $\eP$ up to twist-4 accuracy at the LO and evaluating twist-2 and twist-3 contributions, including the gluonic twist-2 part, at the NLO. 
The $B, B_s$ and $D, D_s$ decay constants can also be calculated in the same scheme using the standard sum rule expression from Ref.~\cite{Jamin:2001fw} at $\mathcal{O}(\alpha_s, m_s^2)$.
However, since these results differ from the most recent, very precise lattice results by about $\pm 10\%$, depending on the heavy meson, we have decided to use the lattice results here, in contrast to our previous work \cite{Duplancic:2015zna}. 

\subsection{Numerical analysis of the form factors}
We closely follow the analysis developed in Ref.~\cite{Leljak:2021vte}.
The varied input parameters of our sum rules are summarised in \cref{tab:inputs}, the fixed parameters are given in \cref{app:fixed_params}.
The threshold parameters are obtained using the so-called daughter sum rule
\begin{equation}\label{eq:daughter-sum-rule}
    \left. m_H(q^2) \right|_\mathrm{LCSR} = \frac{\int_0^{s_0} s \rho(s, q^2) e^{-s / M^2} ds}{\int_0^{s_0} \rho(s, q^2) e^{-s / M^2} ds} \,,
\end{equation}
where $\rho$ is the spectral density of the $H\to\eta^{(\prime)}$ form factor and $H$ stands for $B, B_s, D$ or $D_s$.
We require that the LCSR masses agree with the experimental ones with a precision smaller than $2\%$ by applying a Gaussian penalty to all samples.

\begin{table}[ht]
    \centering
    \begin{tabular}{l ccc}
    \toprule
        & Prior distribution & $B_{(s)} \to \eta^{(\prime)}$ & $D_{(s)} \to \eta^{(\prime)}$ \\
    \midrule
        Borel $M^2$                 & uniform  & $[15, 21]\,\mathrm{GeV}^2$ & $[4, 9]\,\mathrm{GeV}^2$ \\
        threshold $s_0$             & uniform  & $[31, 42]\,\mathrm{GeV}^2$ & $[4, 9]\,\mathrm{GeV}^2$ \\
        $a_{2}^{\pi}(1\,\mathrm{GeV})$ & Gaussian & \multicolumn{2}{c}{$0.17 \pm 0.08$~\cite{Khodjamirian:2011ub}} \\
        $a_{4}^{\pi}(1\,\mathrm{GeV})$ & Gaussian & \multicolumn{2}{c}{$0.06 \pm 0.10$~\cite{Khodjamirian:2011ub}} \\
        $b_2^g(1\,\mathrm{GeV})$      & Gaussian & \multicolumn{2}{c}{$0 \pm 20$~\cite{Ball:2007hb}} \\
        $\phi$                      & Gaussian & \multicolumn{2}{c}{$(39.3 \pm 1)^\circ$~\cite{Feldmann:1998vh}} \\
        $\mu$                       & uniform  & $[0.75, 1.25] \sqrt{M_{B_{(s)}}^2 - m_b^2}$ & $[0.75, 1.25] \sqrt{M_{D_{(s)}}^2 - m_c^2}$ \\
    \bottomrule
    \end{tabular}
    \caption{Priors models and ranges used to evaluate the sum rules.}
    \label{tab:inputs}
\end{table}

There are three main sources of systematic uncertainties: the dependence of the sum rule on the threshold $s_0$, the scale $\mu$ at which the parameters are evaluated and the mixing angle of the $\eta-\eta'$ system.
\begin{itemize}
    \item Following Ref.~\cite{Leljak:2021vte}, we consider a systematic uncertainty due to the sum rule threshold by allowing $s_0$ to vary with $q^2$.
    This second model is physically motivated by the $q^2$ dependence of the spectral density in \cref{eq:daughter-sum-rule}.
    We, therefore, set $s_0(q^2) = s_0 + s_0' \, q^2$ and vary $s_0'$ uniformly in $[0, 0.2]$ for all form factors.
    This model results in a $q^2$-dependent increase of the form factors, and we consider the relative shift as a 100\% correlated systematic uncertainty.
    For $B$ decays, the largest shift is obtained for $F_{B\to\eta}(q^2 = 5\,\mathrm{GeV}^2)$ and amounts to a relative uncertainty of 4\%, in agreement with the findings of Ref.~\cite{Leljak:2021vte}.
    For $D$ decays, the sum rules are less stable and the largest shift is obtained for $F_{D\to\eta'}(q^2 = 5\,\mathrm{GeV}^2)$ with a relative uncertainty of 9\%.

    \item For each form factor, we estimate the systematic uncertainty due to the $\mu$-scale variation by varying this scale by 25\% above and below its nominal value $\mu = \sqrt{M_{B_{(s)}}^2 - m_b^2}$ or $\mu = \sqrt{M_{D_{(s)}}^2 - m_c^2}$.
    For all form factors, the most significant effect is obtained when lowering the scale by 25\%, resulting in an increase of the form factor's value by at most 4\%.
    Although this uncertainty is, in principle, correlated amongst the different form factors and $q^2$ values, we consider an uncorrelated uncertainty of 4\% for each value.

    \item For most transitions, the largest source of systematic uncertainties is due to the $\eta - \eta'$ mixing.
    As described above, we model the mixing with a single angle $\phi$ that we assume Gaussian distributed.
    We neglect correlations between $\phi$ and the other parameters and vary it independently.
\end{itemize}

We provide the values obtained for the form factors at $q^2 = 0$ for comparison.
For $B, B_s$ decays, we get
\begin{equation}
\begin{aligned}
    f^{+,0}_{B\eta}(0) = 0.185 \pm 0.007_\mu \pm 0.007_{s_0} \pm 0.049_{h_q,\phi} \pm 0.008_\mathrm{rest} &= 0.185 \pm 0.050 \\
    f^{+,0}_{B\eta'}(0) = 0.144 \pm 0.006_\mu \pm 0.006_{s_0} \pm 0.045_{h_q,\phi} \pm 0.017_\mathrm{rest} &= 0.144 \pm 0.049 \\
    f^T_{B\eta}(0) = 0.191 \pm 0.008_\mu \pm 0.007_{s_0} \pm 0.050_{h_q,\phi} \pm 0.008_\mathrm{rest} &= 0.191 \pm 0.052 \\
    f^T_{B\eta'}(0) = 0.158 \pm 0.006_\mu \pm 0.006_{s_0} \pm 0.049_{h_q,\phi} \pm 0.014_\mathrm{rest} &= 0.158 \pm 0.052
\end{aligned}
\end{equation}
\begin{equation}
\begin{aligned}
    |f^{+,0}_{B_s\eta}(0)| = 0.245 \pm 0.010_\mu \pm 0.005_{s_0} \pm 0.003_{h_q,\phi} \pm 0.009_\mathrm{rest} &= 0.245 \pm 0.015 \\
    f^{+,0}_{B_s\eta'}(0) = 0.277 \pm 0.011_\mu \pm 0.006_{s_0} \pm 0.008_{h_q,\phi} \pm 0.021_\mathrm{rest} &= 0.277 \pm 0.025 \\
    |f^T_{B_s\eta}(0)| = 0.254 \pm 0.010_\mu \pm 0.005_{s_0} \pm 0.003_{h_q,\phi} \pm 0.009_\mathrm{rest} &= 0.254 \pm 0.015 \\
    f^T_{B_s\eta'}(0) = 0.309 \pm 0.012_\mu \pm 0.007_{s_0} \pm 0.008_{h_q,\phi} \pm 0.017_\mathrm{rest} &= 0.309 \pm 0.024.
\end{aligned}
\end{equation}
For $D, D_s$ decays, we obtain 
\begin{equation}
\begin{aligned}
    f^{+,0}_{D\eta}(0) = 0.380 \pm 0.015_\mu \pm 0.010_{s_0} \pm 0.129_{h_q,\phi} \pm 0.010_\mathrm{rest} &= 0.380 \pm 0.130 \\
    f^{+,0}_{D\eta'}(0) = 0.286 \pm 0.011_\mu \pm 0.021_{s_0} \pm 0.111_{h_q,\phi} \pm 0.030_\mathrm{rest} &= 0.286 \pm 0.118 \\
    f^T_{D\eta}(0) = 0.380 \pm 0.015_\mu \pm 0.012_{s_0} \pm 0.129_{h_q,\phi} \pm 0.014_\mathrm{rest} &= 0.380 \pm 0.131 \\
    f^T_{D\eta'}(0) = 0.342 \pm 0.014_\mu \pm 0.031_{s_0} \pm 0.133_{h_q,\phi} \pm 0.032_\mathrm{rest} &= 0.342 \pm 0.141 \\
\end{aligned}
\end{equation}
\begin{equation}
\begin{aligned}
    |f^{+,0}_{D_s\eta}(0)| = 0.467 \pm 0.019_\mu \pm 0.003_{s_0} \pm 0.005_{h_q,\phi} \pm 0.010_\mathrm{rest} &= 0.467 \pm 0.022 \\
    f^{+,0}_{D_s\eta'}(0) = 0.501 \pm 0.020_\mu \pm 0.044_{s_0} \pm 0.008_{h_q,\phi} \pm 0.024_\mathrm{rest} &= 0.501 \pm 0.054\\
    |f^T_{D_s\eta}(0)| = 0.456 \pm 0.018_\mu \pm 0.006_{s_0} \pm 0.005_{h_q,\phi} \pm 0.015_\mathrm{rest} &= 0.456 \pm 0.025 \\
    f^T_{D_s\eta'}(0) = 0.578 \pm 0.023_\mu \pm 0.042_{s_0} \pm 0.010_{h_q,\phi} \pm 0.031_\mathrm{rest} &= 0.578 \pm 0.058.
\end{aligned}
\end{equation}
The uncertainties are separated between the scale dependence, the threshold modelling, the uncertainty due to $h_q$ and the $\eta - \eta'$ mixing, and the other parametric uncertainties.
We remind that we use lattice QCD inputs \cite{FlavourLatticeAveragingGroupFLAG:2024oxs} for the decay constants.
These results are therefore in perfect agreement with those of Ref.~\cite{Duplancic:2015zna}, where decay constants were evaluated using two-point sum rules.

In anticipation of our phenomenological analysis of~\cref{sec:CKM_extraction}, we observe that our results are also in excellent agreement with the values obtained at $q^2 = 0$ by the experimental collaborations for $D\to\eta$~\cite{BESIII:2025hjc}, $D\to\eta^\prime$~\cite{BESIII:2024njj}, and $D_s\to\eta^{(\prime)}$~\cite{BESIII:2023gbn} transitions.

Our results agree with existing similar LCSR calculations \cite{Offen:2013nma,Hu:2021zmy,Hu:2023pdl,Zhang:2025yeu} within uncertainties.
A first lattice calculation of $D_s \to \eta, \eta'$ scalar form factors also exists at zero momentum transfer with a single lattice spacing \cite{Bali:2014pva}.
In this approximation, they obtain $|f^0_{D_s \eta}(0)| = 0.542(13)$ and $|f^0_{D_s \eP}(0)| = 0.404(25)$ (at $M_{\pi} = 370$ MeV).
This implies $|f^0_{D_s \eta}(0)|/|f^0_{D_s \eP}(0)| > 1$, which shows tension with estimates from the LCSRs, other models \cite{Wei:2009nc,Ivanov:2019nqd}, and the latest experimental results \cite{BESIII:2023ajr}.

\subsection{Form factors in the full $q^2$ range}\label{sec:extrapolation}

We now use these updated predictions to present a set of phenomenology applications.
To extend the validity of the form factors to large $q^2$ values, beyond the LCSR reach, we fit the form factors to a BSZ expansion~\cite{Bharucha:2015bzk}
\begin{equation}
    \label{eq:SSE}
    f(q^2) = \frac{1}{1 - q^2/m_R^2} \sum_{n=0}^N \alpha_n^f \, \left(z(q^2) - z(0)\right)^n,
\end{equation}
where
\begin{equation}
    z(q^2) = \frac{\sqrt{t_+ - q^2} - \sqrt{t_+ - t_0}}{\sqrt{t_+ - q^2} + \sqrt{t_+ - t_0}},
\end{equation}
$t_\pm$ is the squared sum (respectively difference) of the initial and final state mesons, and $t_0$ is a constant set to its optimal value $t_0 = t_+ \left( 1 - \sqrt{1 - \frac{t_-}{t_+}}\right)$.
Furthermore, $m_R$ is the mass of the lowest-lying resonance with quantum numbers compatible with the form-factor $f$.
The resonance masses are provided in \cref{tab:resonances}, in \cref{app:fixed_params}.
Since the form factors are only constrained by low-$q^2$ data, extrapolation effects could be large.
To estimate those, we apply a complete series expansion (SE)~\cite{Boyd:1994tt,Bharucha:2010im}, following the implementation suggested in Ref.~\cite{Gubernari:2023puw}.
The effects of the dispersive bounds on the extrapolation uncertainties are found to be small in the SE, and the SE uncertainties are correctly reproduced by setting $N=2$ in \cref{eq:SSE}.
We keep a more detailed analysis of the extrapolation uncertainties for future work.

\begin{figure}[ht]
    \centering
    \includegraphics[width=0.49\linewidth]{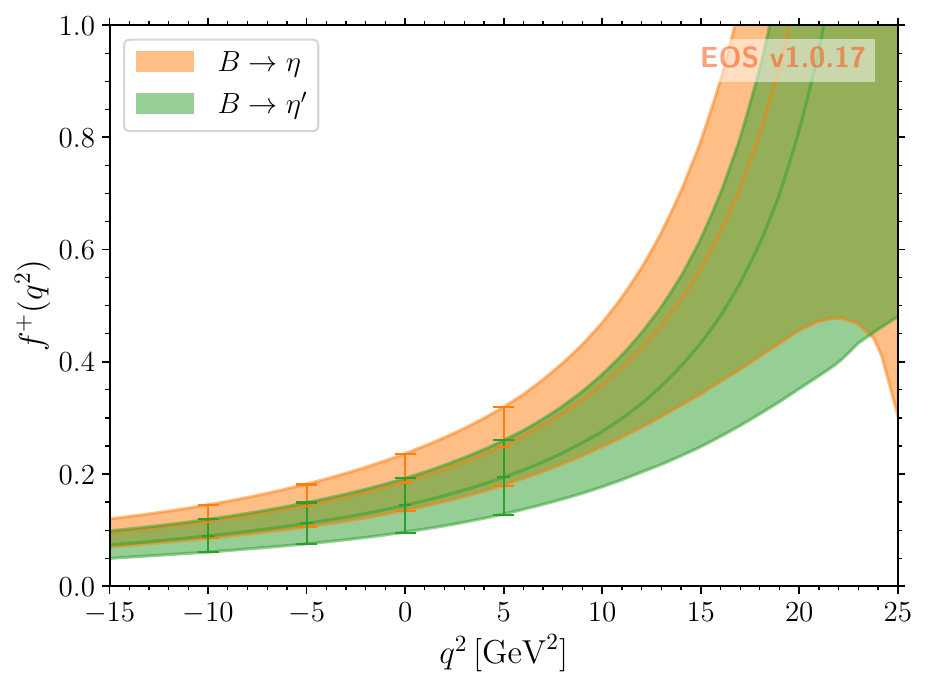}
    \includegraphics[width=0.49\linewidth]{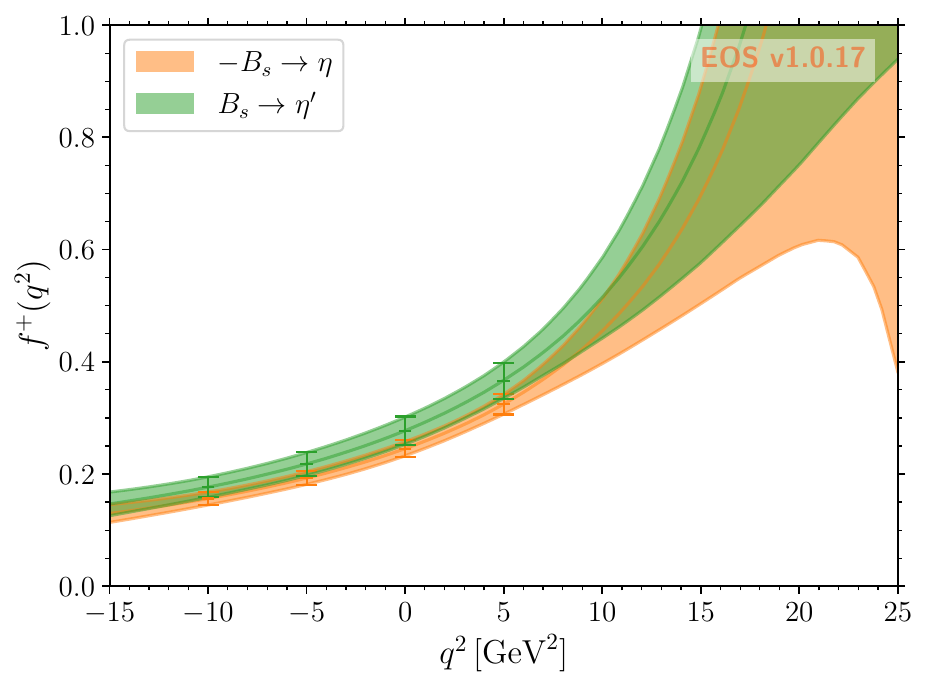}\\
    \includegraphics[width=0.49\linewidth]{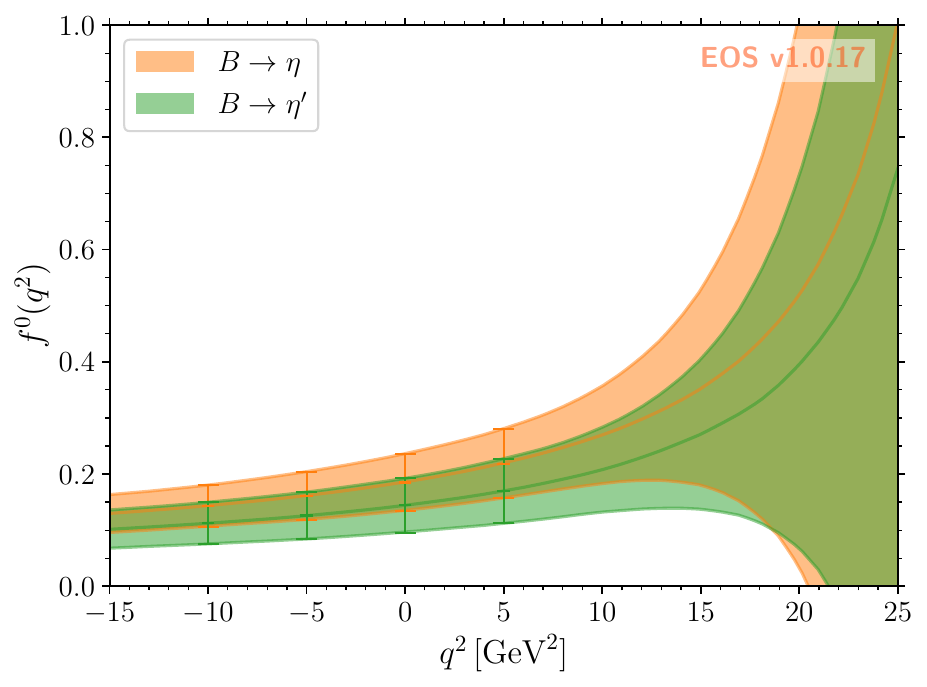}
    \includegraphics[width=0.49\linewidth]{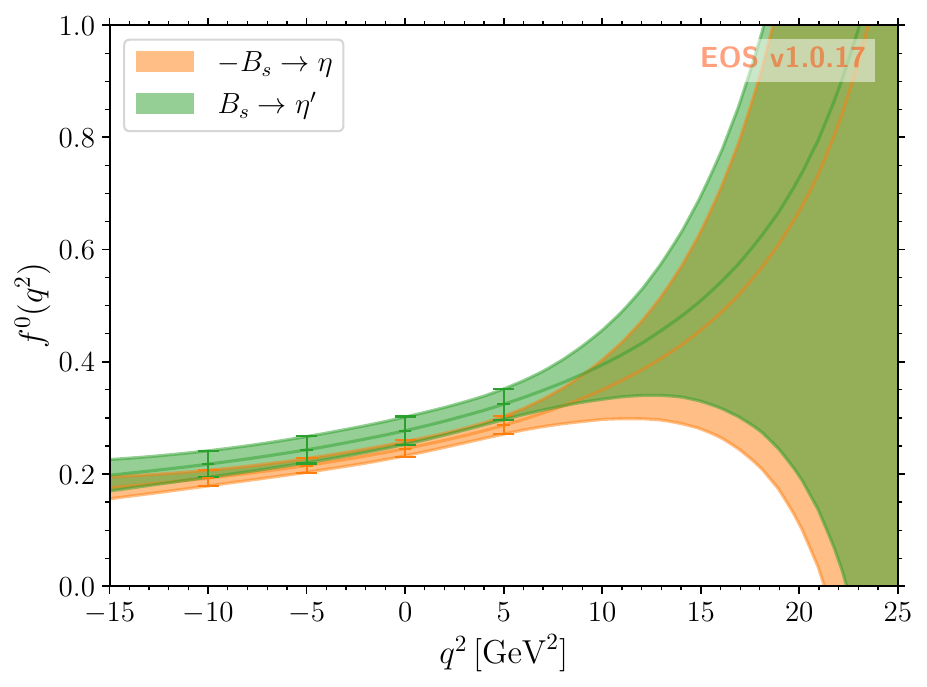}\\
    \includegraphics[width=0.49\linewidth]{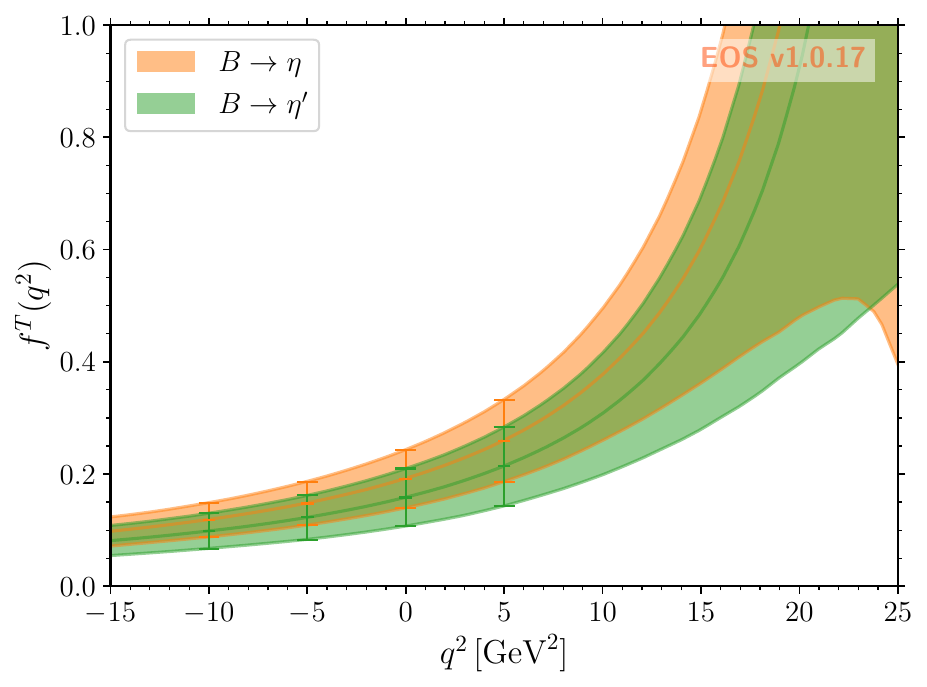}
    \includegraphics[width=0.49\linewidth]{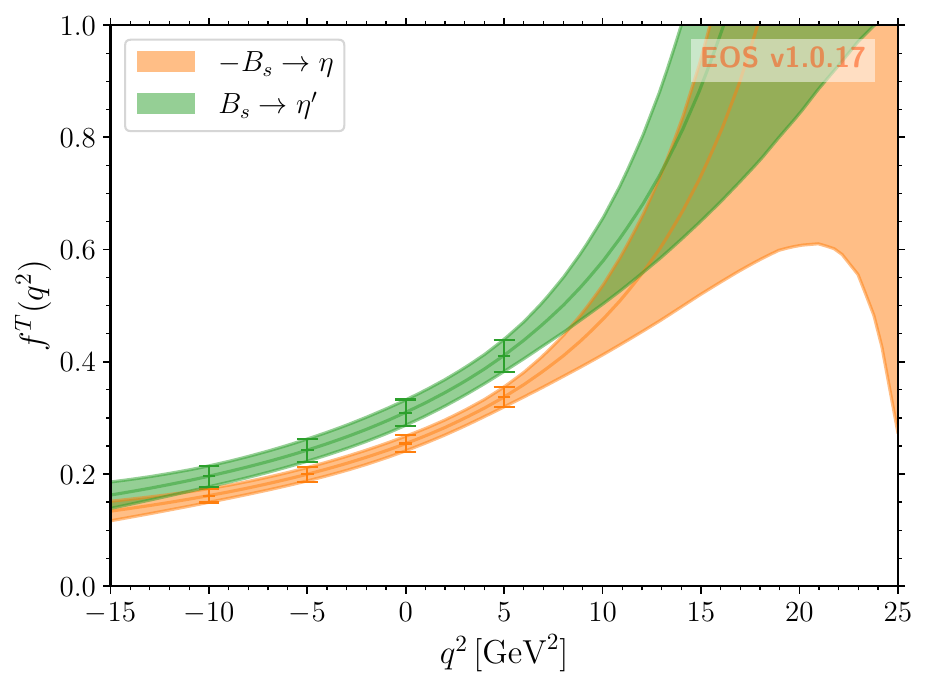}
    \caption{Summary plots of our results for $B \to \eT, \eP$ and $B_s \to \eta, \eP$ form factors.
    The error bars correspond to the LCSR results.
    The shaded areas are the $1\sigma$ uncertainty bands of our extrapolation~\cref{eq:SSE} with a truncation order $N=2$.}
    \label{fig:B_ffs}
\end{figure}

\begin{figure}[ht]
    \centering
    \includegraphics[width=0.49\linewidth]{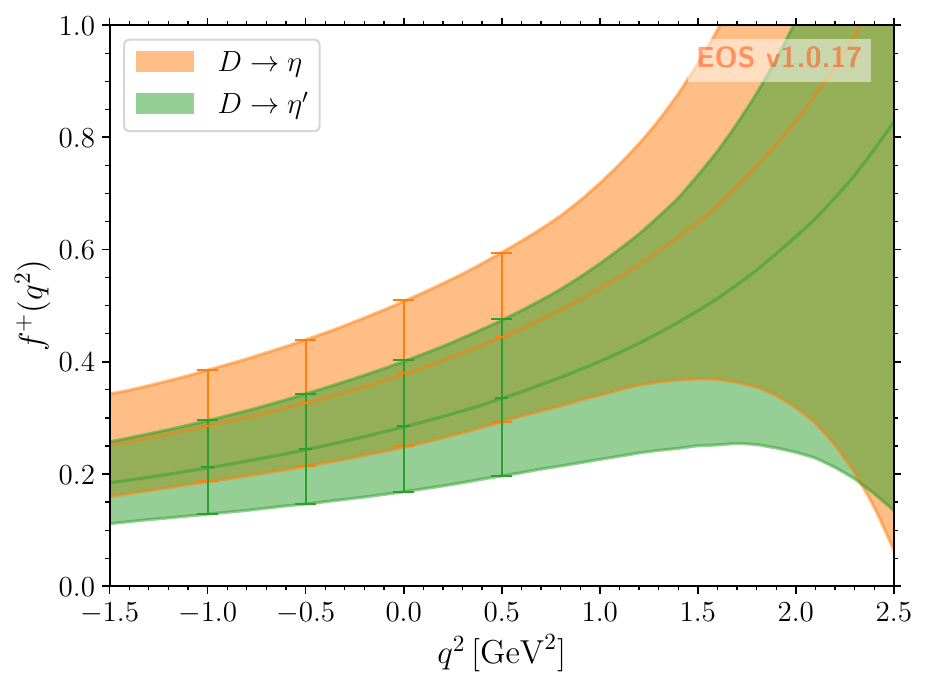}
    \includegraphics[width=0.49\linewidth]{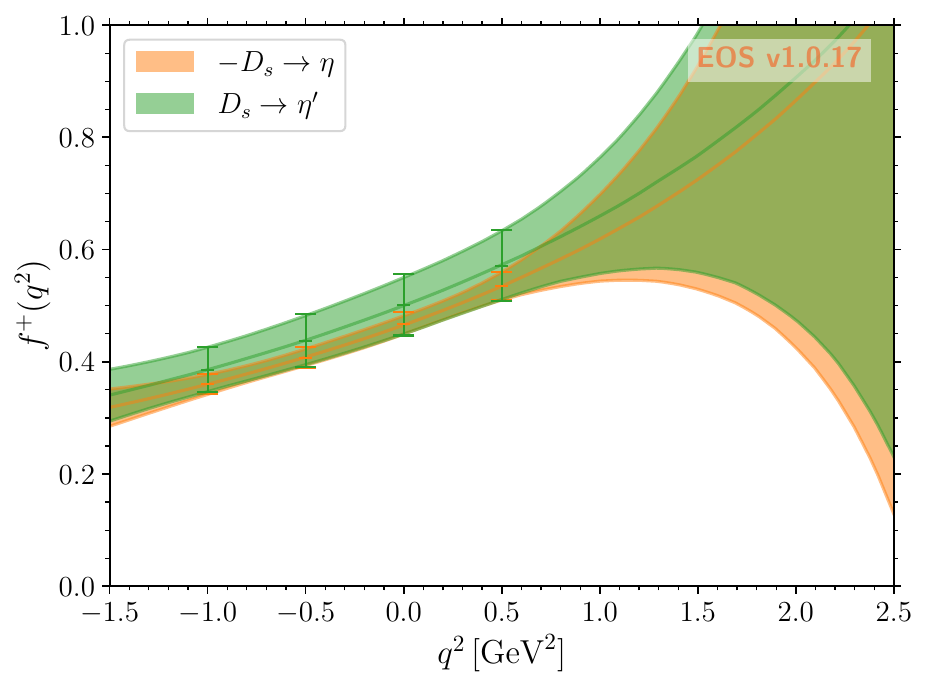}\\
    \includegraphics[width=0.49\linewidth]{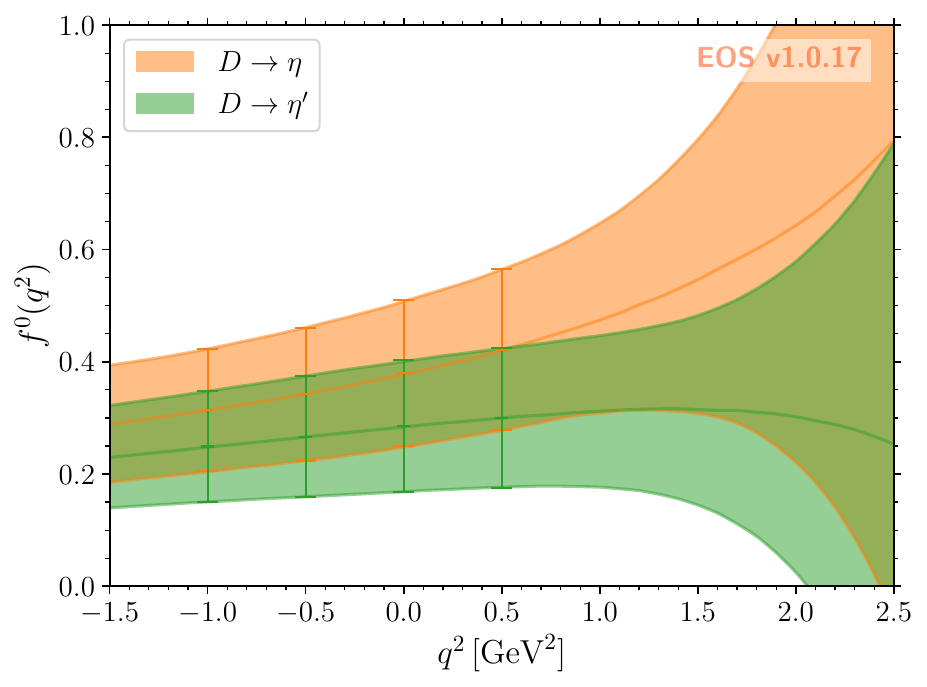}
    \includegraphics[width=0.49\linewidth]{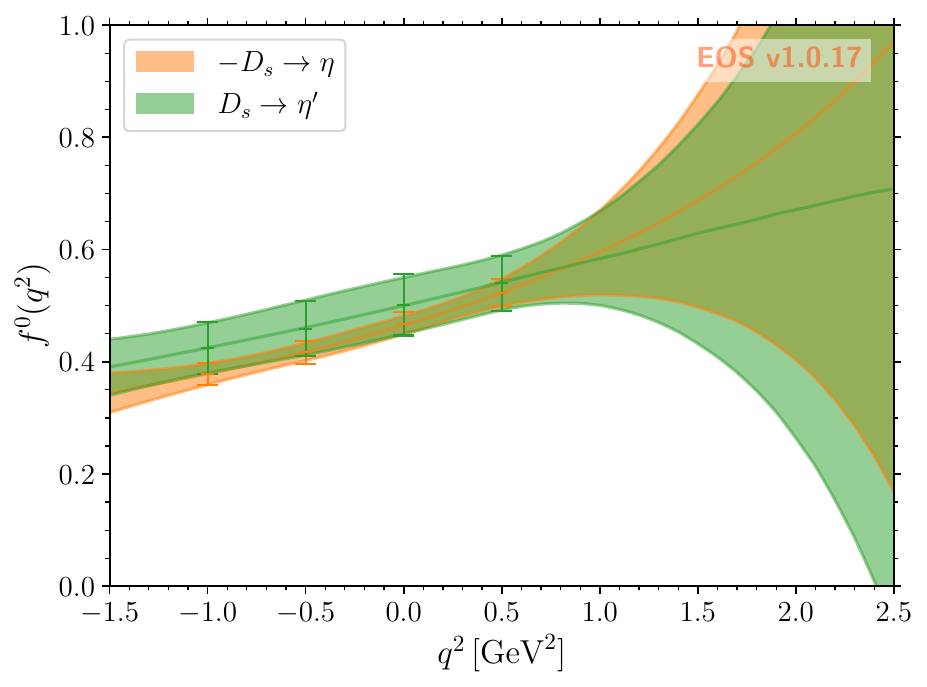}\\
    \includegraphics[width=0.49\linewidth]{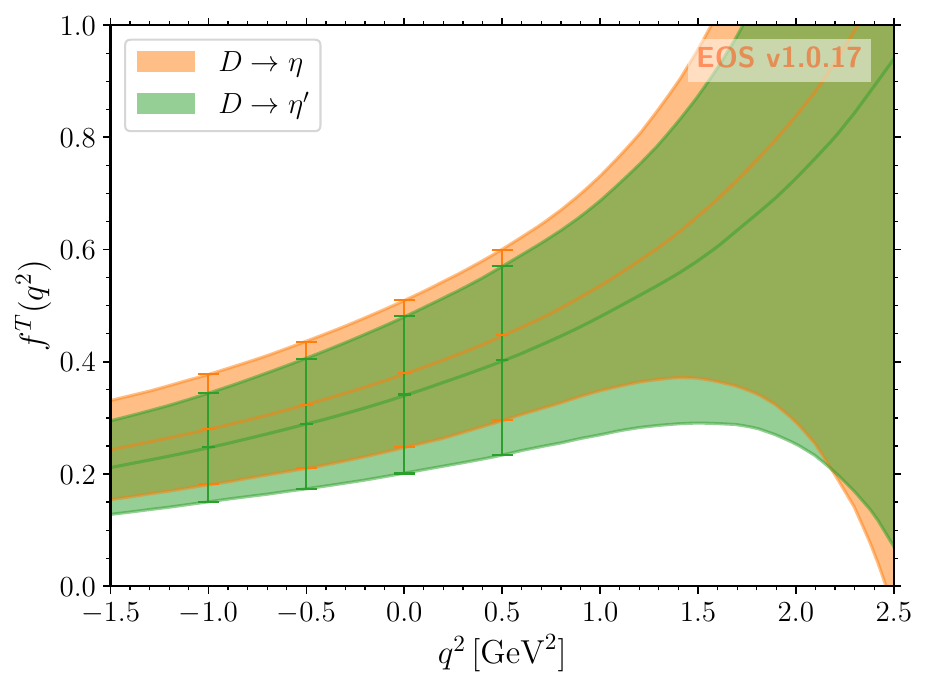}
    \includegraphics[width=0.49\linewidth]{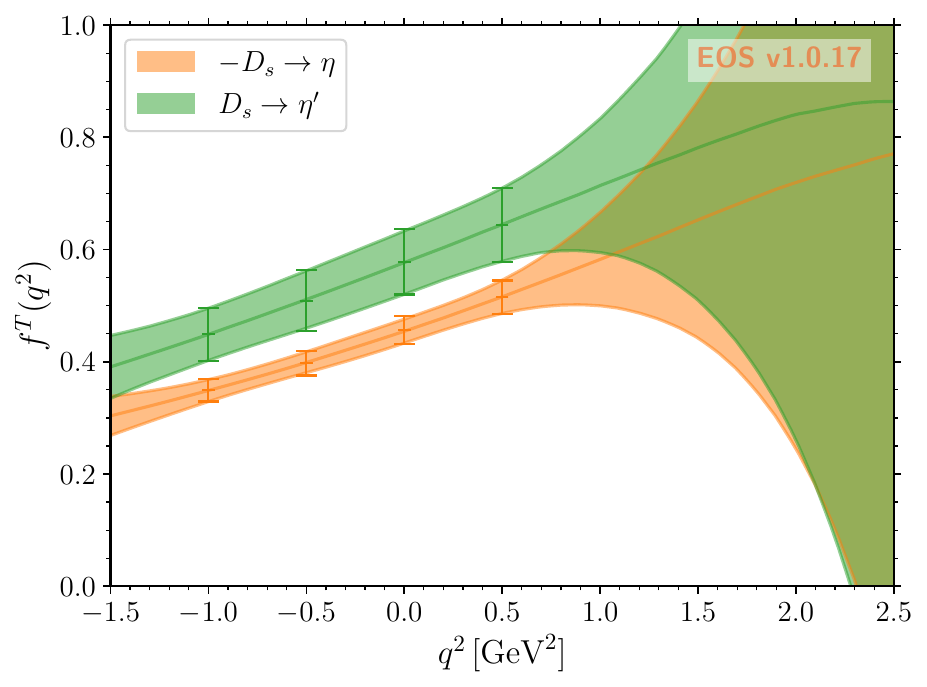}
    \caption{Summary plots of our results for $D \to \eT, \eP$ and $D_s \to \eta, \eP$ form factors.
    The error bars correspond to the LCSR results.
    The shaded areas are the $1\sigma$ uncertainty bands of our extrapolation~\cref{eq:SSE} with a truncation order $N=2$.}
    \label{fig:D_ffs}
\end{figure}

The numerical analysis is performed using the EOS software~\cite{EOSAuthors:2021xpv} version v1.0.17~\cite{EOS:v1.0.17}.
The code used to run it and all our results are available in the analysis repository~\cite{EOS-DATA-2025-04}.
All the fits have 3 degrees of freedom (11 LCSR constraints for 8 parameters) and show excellent $p$-values larger than $99.8\%$.
The posterior distributions are provided in \cref{app:z_coeff} and the resulting form factors are shown in \cref{fig:B_ffs,fig:D_ffs}.
As expected, the form factors are well described in the low-$q^2$ region where LCSR constraints are obtained, but the extrapolation uncertainties increase quadratically with $q^2$.

The $f^{+,0}_{D_s\eta^(\prime)}(q^2)$ form factors have also been extracted directly from semileptonic decays in Ref.~\cite{BESIII:2023gbn}.
Since the corresponding numerical values are not provided in the publication, we can only verify that our results show visually very good agreement with their Fig.3.

\subsection{Comparison with the lattice $B_s \to \eta_s$ data}
\begin{figure}[ht]
    \centering
    \includegraphics[width=0.49\linewidth]{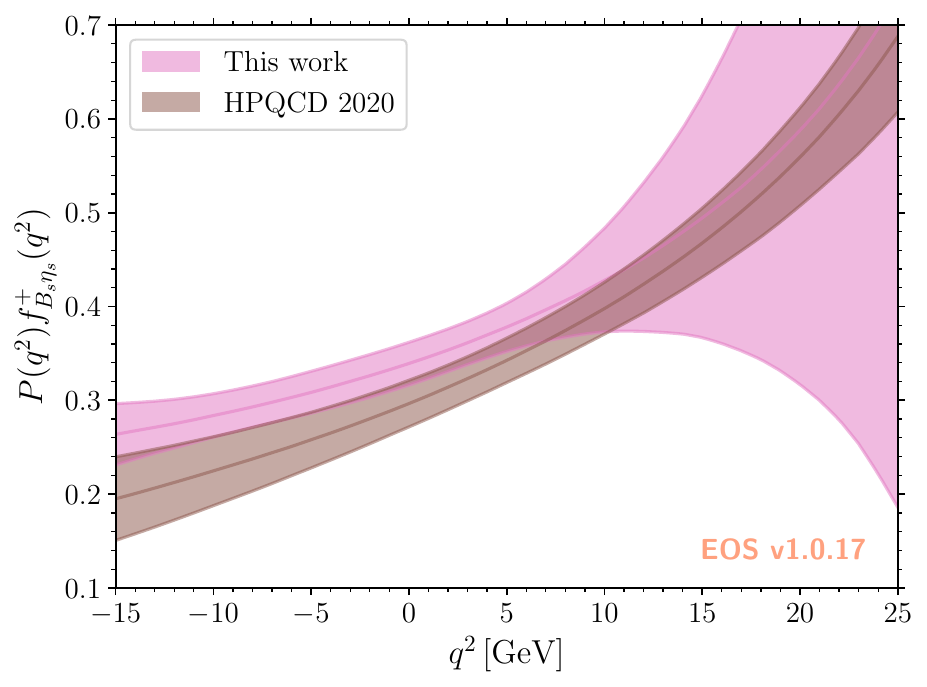}
    \includegraphics[width=0.49\linewidth]{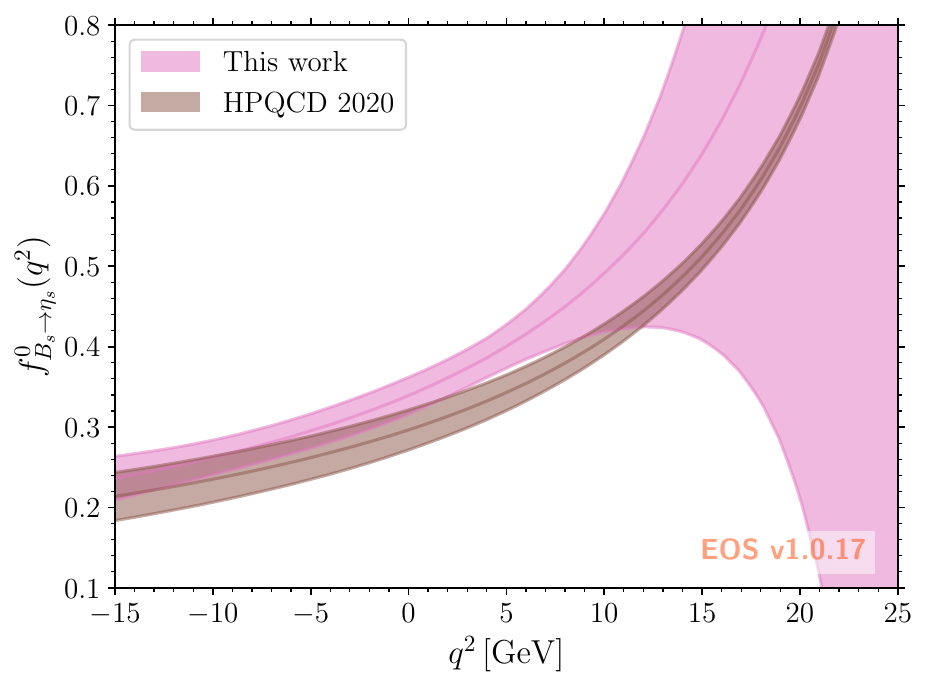}
    \caption{Comparison between our results for unphysical $B_s \to \eta_s$ form factors and the LQCD estimates of Ref.~\cite{Parrott:2020vbe}.}
    \label{fig:BsToEtasvsLQCD}
\end{figure}

$B_s \to \eta_s$ form factors where evaluated on the lattice in Refs.~\cite{Parrott:2020vbe, Bouchard:2014ypa}, where the following results were obtained
\begin{align*}
    f^{0,+}_{B_s\eta_s, \mathrm{LQCD}}(0) & = 0.296(25)\,, \\
    f^+_{B_s\eta_s, \mathrm{LQCD}}(q^2_{\rm max}) & = 2.58(28)\,, \\
    f^0_{B_s\eta_s, \mathrm{LQCD}}(q^2_{\rm max}) & = 0.808(15)(27)\,.
\end{align*}
To check these numbers, we have adapted our calculations of $f^+(q^2)$ and $f^0(q^2)$ form factors to the non-physical contribution from the $B_s \to \eta_s$ transition.
The estimation is straightforward, as it reduces to neglecting the $\eta_q$ contributions and replacing $h_s$ by its chiral perturbation theory approximation from \cref{eq:hq-approx}.
We have also used a fixed mass $m_{\eta_s} = 0.6885$ GeV~\cite{Dowdall:2013rya} in the LCSR estimation and the extrapolation to high $q^2$ as described in \cref{sec:extrapolation}.
With these modifications, our results read
\begin{align*}
    f^{+,0}_{B_s\eta_s}(0) & = 0.339 \pm 0.014_\mu \pm 0.013_{s_0} \pm 0.017_\mathrm{rest} = 0.339 \pm 0.025\,, \\
    f^+_{B_s\eta_s}(q^2_{\rm max}) & = 2.5 \pm 1.3\,, \\
    f^0_{B_s\eta_s}(q^2_{\rm max}) & = 1.1 \pm 1.0\,.
\end{align*}
We provide a comparison of these form factors with the lattice results in \cref{fig:BsToEtasvsLQCD}.

Despite larger values at $q^2=0$ and slightly different $q^2$ dependence, the agreement between our LCSR predictions and the lattice results is overall very good.

We also note that our numbers are very close to the physical form factors $B_s \to \eta$, when multiplied by $-\sin(\phi)$.
This is expected, since the anomaly and gluonic NLO contributions are small for the $B_s \to \eta$ form factors.

\section{SM predictions}

We now provide updated SM predictions for the main experimental observables that rely on our form factors.
Our experimental inputs (masses, CKM elements, \dots) are given in \cref{app:fixed_params}. 
Most of our predictions agree with the experimental measurements within uncertainties.
Small tensions in the normalisations are discussed in \cref{sec:CKM_extraction}, where CKM elements are extracted from experimental data.

The predictions for the $B_{(s)}, D_{(s)}\to \eta^{(\prime)} \ell \nu$ branching ratios are as follows:
\begin{gather}
    \mathcal{B}(B^-\to\eta\ell^-\bar\nu_\ell) = 6.5^{+4.7}_{-3.3} \times 10^{-5} \nonumber \\
    \mathcal{B}(B^-\to\eta'\ell^-\bar\nu_\ell) = 2.7^{+2.2}_{-1.5} \times 10^{-5} \,,
\end{gather}
\begin{eqnarray}
    \mathcal{B}(D^-\to\eta e^-\bar\nu_e) &=& 1.14^{+0.89}_{-0.64} \times 10^{-3} \qquad \mathcal{B}(D^-\to\eta\mu^-\bar\nu_\mu) = 1.12^{+0.88}_{-0.63} \times 10^{-3} \nonumber \\
    \mathcal{B}(D^-\to\eta'e^-\bar\nu_e) &=& 1.46^{+1.45}_{-0.95} \times 10^{-4} \qquad \mathcal{B}(D^-\to\eta'\mu^-\bar\nu_\mu) = 1.37^{+1.36}_{-0.89} \times 10^{-4} \nonumber \\
    \nonumber \\
    \mathcal{B}(D_s^-\to\eta e^-\bar\nu_e) &=& 2.15^{+0.37}_{-0.31} \times 10^{-2} \qquad \mathcal{B}(D_s^-\to\eta\mu^-\bar\nu_\mu) = 2.11^{+0.36}_{-0.30} \times 10^{-2} \\
    \mathcal{B}(D_s^-\to\eta' e^-\bar\nu_e) &=& 6.5^{+1.4}_{-1.3} \times 10^{-3} \,\;\;\qquad \mathcal{B}(D_s^-\to\eta'\mu^-\bar\nu_\mu) = 6.2^{+1.3}_{-1.2} \times 10^{-3} \,, \nonumber
\end{eqnarray}
where $\ell = e, \mu$ is used when no significant difference is found between the two modes, as expected from the lepton flavour universality of the SM and measured in experiments.
Departures from LFU have been tested with leptonic ratios for which we can provide updated predictions.
\begin{eqnarray}
    \frac{\mathcal{B}(B^-\to\eta \mu^-\bar\nu_\mu)}{\mathcal{B}(B^-\to\eta e^-\bar\nu_e)} &=& 0.9997^{+0.0017}_{-0.0010} \quad\qquad \frac{\mathcal{B}(B^-\to\eta'\mu^-\bar\nu_\mu)}{\mathcal{B}(B^-\to\eta'e^-\bar\nu_e)} = 0.9993^{+0.0011}_{-0.0008} \nonumber \\
    \frac{\mathcal{B}(D^-\to\eta \mu^-\bar\nu_\mu)}{\mathcal{B}(D^-\to\eta e^-\bar\nu_e)} &=& 0.9908^{+0.0082}_{-0.0065} \quad\qquad \frac{\mathcal{B}(D^-\to\eta' \mu^-\bar\nu_\mu)}{\mathcal{B}(D^-\to\eta'e^-\bar\nu_e)} = 0.9686^{+0.0038}_{-0.0038} \\
    \frac{\mathcal{B}(D_s^-\to\eta \mu^-\bar\nu_\mu)}{\mathcal{B}(D_s^-\to\eta e^-\bar\nu_e)} &=& 0.9948^{+0.0092}_{-0.0073} \quad\qquad \frac{\mathcal{B}(D_s^-\to\eta'\mu^-\bar\nu_\mu)}{\mathcal{B}(D_s^-\to\eta'e^-\bar\nu_e)} = 0.9774^{+0.0053}_{-0.0045} \,. \nonumber
\end{eqnarray}
An excellent level of agreement is observed between the theoretical predictions and the experimental determination of the ratios for $D, D_s \to \eta,\eta'$ decays provided in \cite{BESIII:2023gbn, BESIII:2024njj, BESIII:2025hjc}.

We also predict the following ratios of branching ratios, which benefit from a partial cancellation of theory uncertainties,
\begin{gather}
    \frac{\mathcal{B}(B^-\to\eta'\ell^-\bar\nu_\ell)}{\mathcal{B}(B^-\to\eta\ell^-\bar\nu_\ell)} = 0.43^{+0.20}_{-0.16} \nonumber \\
    \frac{\mathcal{B}(D^-\to\eta' e^-\bar\nu_e)}{\mathcal{B}(D^-\to\eta e^-\bar\nu_e)} = 0.104^{+0.049}_{-0.048} \qquad \frac{\mathcal{B}(D^-\to\eta'\mu^-\bar\nu_\mu)}{\mathcal{B}(D^-\to\eta\mu^-\bar\nu_\mu)} = 0.100^{+0.047}_{-0.046} \\
    \nonumber \\
    \frac{\mathcal{B}(D_s^-\to\eta' e^-\bar\nu_e)}{\mathcal{B}(D_s^-\to\eta e^-\bar\nu_e)} = 0.346^{+0.103}_{-0.084} \qquad \frac{\mathcal{B}(D_s^-\to\eta'\mu^-\bar\nu_\mu)}{\mathcal{B}(D_s^-\to\eta\mu^-\bar\nu_\mu)} = 0.335^{+0.100}_{-0.080} \,. \nonumber
\end{gather}

Finally, we provide SM predictions for the leptonic forward-backwards asymmetries
\begin{eqnarray}
    \mathcal{A}_\mathrm{FB}(B^-\to\eta e^-\bar\nu_e) &=& -2.038^{+0.030}_{-0.034} \times 10^{-6} \qquad \mathcal{A}_\mathrm{FB}(B^-\to\eta\mu^-\bar\nu_\mu) = -2.788^{+0.033}_{-0.036} \times 10^{-2} 
    \nonumber \\
    \mathcal{A}_\mathrm{FB}(B^-\to\eta'e^-\bar\nu_e) &=& -0.440^{+0.072}_{-0.090} \times 10^{-6} \qquad \mathcal{A}_\mathrm{FB}(B^-\to\eta'\mu^-\bar\nu_\mu) = -0.76^{+0.11}_{-0.13} \times 10^{-2} \,,
    \nonumber  \\
\end{eqnarray}
\begin{eqnarray}
    \mathcal{A}_\mathrm{FB}(D^-\to\eta e^-\bar\nu_e) &=& -5.20^{+0.68}_{-0.71} \times 10^{-6} \,\;\qquad \mathcal{A}_\mathrm{FB}(D^-\to\eta\mu^-\bar\nu_\mu) = -5.90^{+0.60}_{-0.61} \times 10^{-2} \nonumber \\
    \mathcal{A}_\mathrm{FB}(D^-\to\eta'e^-\bar\nu_e) &=& -10.96^{+ 0.42}_{-0.47} \times 10^{-6} \qquad \mathcal{A}_\mathrm{FB}(D^-\to\eta'\mu^-\bar\nu_\mu) = -9.88^{+0.27}_{-0.30} \times 10^{-2} \nonumber \\  \nonumber   \\
    \mathcal{A}_\mathrm{FB}(D_s^-\to\eta e^-\bar\nu_e) &=& -4.65^{+0.67}_{-0.74} \times 10^{-6} \qquad \mathcal{A}_\mathrm{FB}(D_s^-\to\eta\mu^-\bar\nu_\mu) = -5.48^{+0.64}_{-0.65} \times 10^{-2} \nonumber \\
    \mathcal{A}_\mathrm{FB}(D_s^-\to\eta'e^-\bar\nu_e) &=& -9.16^{+0.43}_{-0.45} \times 10^{-6} \qquad \mathcal{A}_\mathrm{FB}(D_s^-\to\eta'\mu^-\bar\nu_\mu) = -8.79^{+0.31}_{-0.33} \times 10^{-2} \,. \nonumber  \\
\end{eqnarray}
The predicted $\mathcal{A}_{BF}$ for $D^-\to\eta\mu^-\bar\nu_\mu$ and $D^-\to\eta'\mu^-\bar\nu_\mu$ decays are in a very good agreement with the experiment \cite{BESIII:2023gbn}.

We don't provide predictions for the rare $B_s \to \eta^{(\prime)} \ell^+\ell^-$ decays, since these decays receive significant contributions from nonlocal hadronic form factors which are beyond this phenomenology analysis~\cite{Khodjamirian:2010vf,Gubernari:2022hxn}.
However, we provide SM predictions for rare $B_s$ decays to $\eta^{(\prime)}$ mesons and a pair of neutrinos, which might be accessible to future colliders.
We obtain
\begin{equation}
    \mathcal{B}(B_s^0\to\eta\nu\bar\nu) = 2.60_{-0.66}^{+0.89} \times 10^{-6}, \qquad
    \mathcal{B}(B_s^0\to\eta^\prime\nu\bar\nu) = 2.37_{-0.50}^{+0.58} \times 10^{-6},
\end{equation}
where we summed over the three lepton families.
Despite their small BRs, these decays could be useful, once combined with purely leptonic $B_s \to \mu^+\mu^-$ decay, for probing potential effects of new physics in penguin diagrams.
The predicted BRs are in agreement with existing results obtained in different models within uncertainties \cite{Carlucci:2009gr,Azizi:2010rx,Choi:2010zb,Faustov:2013pca}.

\subsection{CKM matrix elements extractions} \label{sec:CKM_extraction}
With our predicted form factors at hand, we are now in a position to check the consistency of our results by extracting the CKM matrix elements, $V_{cd}$, $V_{cs}$ and $|V_{ub}|$.
To do so, we perform combined fits of our LCSR results and the available experimental data, listed in \cref{app:exp_inputs}, for all transitions.
We obtain satisfactory $p$-values for all the fits, denoting a good agreement between our predictions and the available experimental data.
A summary of all the fits is given in \cref{tab:fits} and comparison plots between the results of our fits and the experimental inputs are presented in \cref{app:distribution_results}.

\begin{table}[ht]
    \centering
    \begin{tabular}{lcccc}
        \toprule
        Decay & $\chi^2$ & d.o.f. & $p$-value $[\%]$ & $V_{ij}$\\
        \midrule
        $B\to\eta\ell\nu$    & 0.61 &  7 & 99.8 & $2.84_{-0.69}^{+1.16} \times 10^{-3}$ \\[3pt]
        $D\to\eta\ell\nu$    & 5.57 & 10 & 85.0 & $0.199_{-0.047}^{+0.087}$ \\[3pt]
        $D\to\eta'\ell\nu$   & 11.0 & 10 & 36.0 & $0.32_{-0.11}^{+0.18}$ \\[3pt]
        $D_s\to\eta\ell\nu$  & 6.73 & 18 & 99.2 & $0.977_{-0.035}^{+0.035}$ \\[3pt]
        $D_s\to\eta'\ell\nu$ & 3.46 &  8 & 90.2 & $1.062_{-0.088}^{+0.083}$ \\[3pt]
        \midrule
        $B\to\eta^{(\prime)}\ell\nu$   & 0.78 & 11 & 100  & $2.92_{-0.56}^{+0.80} \times 10^{-3}$ \\[3pt]
        $D\to\eta^{(\prime)}\ell\nu$   & 16.8 & 21 & 72.1 & $0.207_{-0.035}^{+0.045}$ \\[3pt]
        $D_s\to\eta^{(\prime)}\ell\nu$ & 11.0 & 27 & 99.7 & $0.987_{-0.033}^{+0.034}$ \\
        \bottomrule
    \end{tabular}
    \caption{Summary of the fits performed by combining our LCSR results and experimental data while varying the CKM elements.}
    \label{tab:fits}
\end{table}

\begin{figure}[th]
    \centering
    \includegraphics[width=0.49\linewidth]{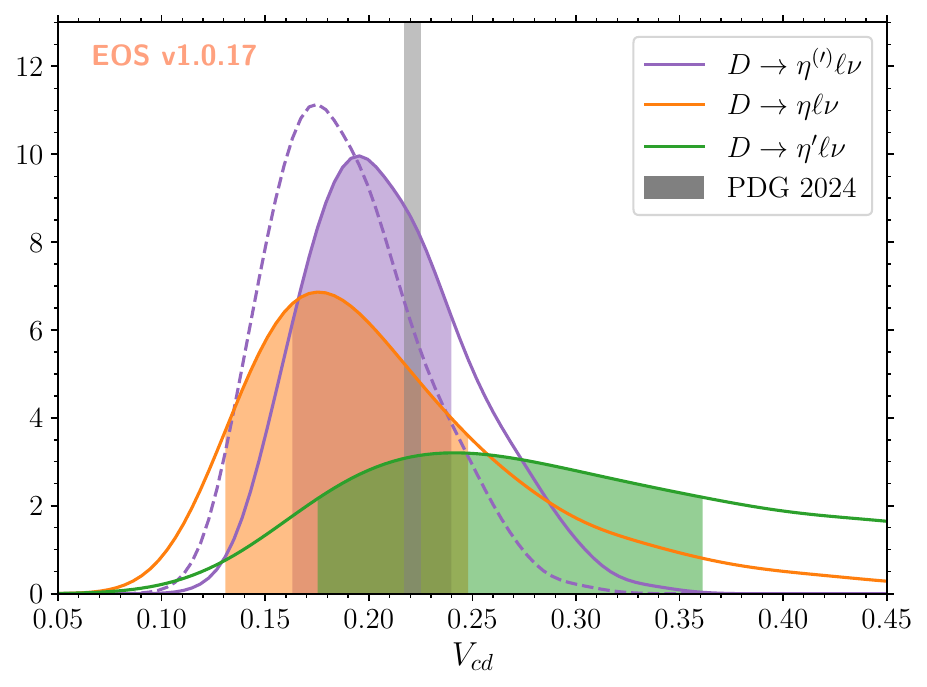}
    \includegraphics[width=0.49\linewidth]{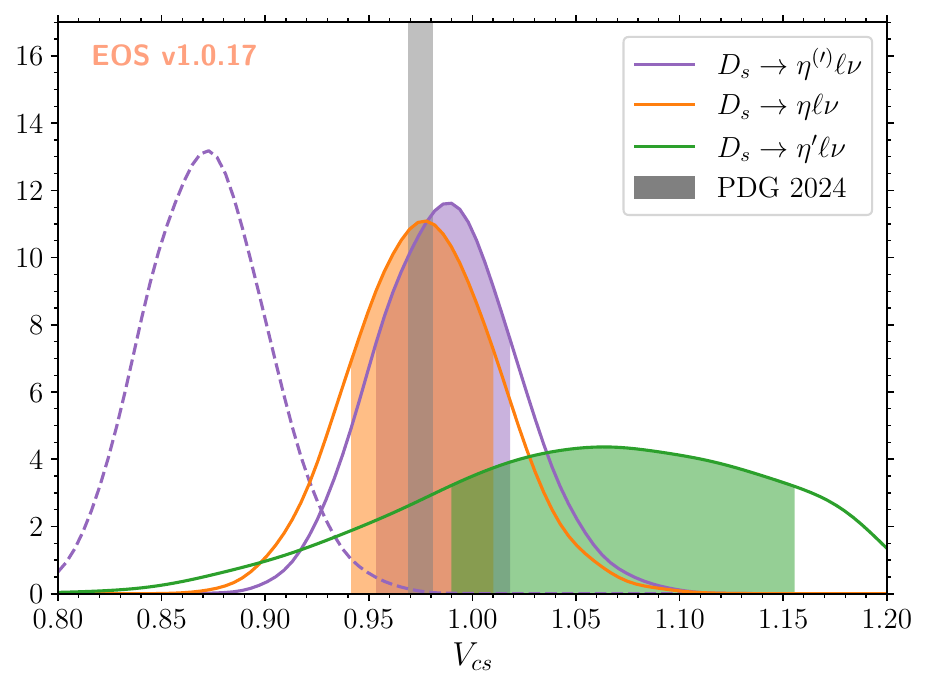}\\
    \includegraphics[width=0.49\linewidth]{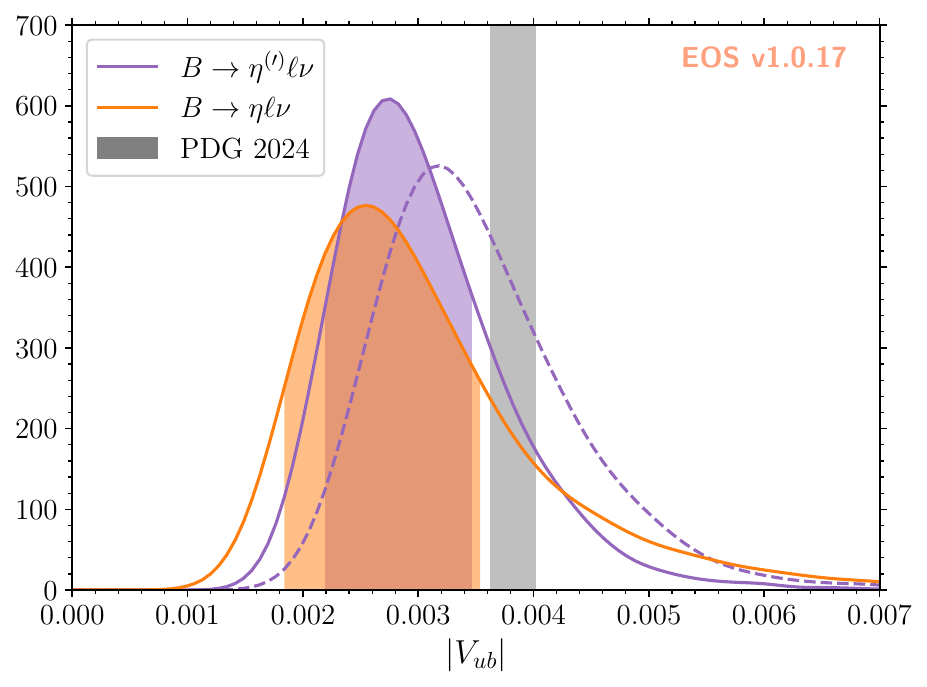}
    \caption{Marginalized posteriors for the CKM parameters $V_{cd}, V_{cs}$ and $|V_{ub}|$ obtained with a Kernel Density Estimate.
    The filled regions correspond to the centred 68\% probability intervals.
    The dashed lines show the total posterior using sum rules instead of lattice inputs for the decay constants.
    The grey line corresponds to the current world averages from Ref.~\cite{ParticleDataGroup:2024cfk}.}
    \label{fig:CKM}
\end{figure}

Focusing just on the extraction of the CKM element, we obtain
\begin{gather}
    V_{cd}   = 0.207_{-0.035}^{+0.045} \,, \qquad
    V_{cs}   = 0.987_{-0.033}^{+0.034} \,, \qquad
    |V_{ub}| = 2.92_{-0.56}^{+0.80} \times 10^{-3}.
\end{gather}
These values are in good agreement with the world averages \cite{ParticleDataGroup:2024cfk,Charles:2004jd,UTfit:2022hsi}, with the largest tension slightly exceeding $1 \sigma$ for $|V_{ub}|$.
The marginalised posterior predictions for the CKM parameters are shown in \cref{fig:CKM}.
The uncertainties we obtain are not much larger than the ones obtained in other exclusive modes, and we look forward to global analyses of all these decays.

For completeness, we overlaid the distributions of \cref{fig:CKM} with those obtained using sum rule inputs for the decay constants (in dashed lines).
In this scenario, we obtain a slightly better agreement for $|V_{ub}|$, but the agreement decreases for $V_{cd}$ and $V_{cs}$. 

\subsection{Discussion of $\eT-\eP$ mixing angle extracted from the data}
\label{sec:phi-dep}

Motivated by the behaviour of the form factor ratios 
\ba
\frac{f_{H\eP}^{+}}{f_{H\eT}^{+}} \sim {\rm tan} \,  \phi \,, \qquad \frac{f_{H_s\eP}^{+}}{f_{H_s\eT}^{+}} \sim -{\rm cot} \, \phi \,,
\label{eq:ctan}
\ea
ratios of branching ratios $\mathcal{B}(H_{(s)} \to \eP \ell\nu) / \mathcal{B}(H_{(s)} \to \eT \ell\nu)$ have been used in the literature to experimentally access the $\eT-\eP$ mixing angle.
For $D \to \eT,\eP\ell \nu$ decays, the BESIII collaboration initially obtained~\cite{BESIII:2018eom} 
\be
\phi = 40 \pm 3 \pm 3^{\circ}\,.
\ee
Including more precise BRs measurements of $D_s \to \eT,\eP\ell \nu$ decays, the precision of the extracted mixing angle is significantly improved and now reads \cite{BESIII:2024njj, BESIII:2019qci}
\be
\phi = 39.8 \pm 0.8 \pm 0.3^{\circ} \,.
\ee
These experimental results nicely agree with the FKS input \cref{eq:FKS} used in the paper.

Having explicit formulas for the form factors in the LCSR framework for $D_{(s)}, B_{(s)} \to \eT, \eP$ decays, we can discuss the validity of this method in our framework.
Due to practicality, we had to stick to a single mixing-angle picture.
As already mentioned, however, the dynamics of various twist contributions to the form factors spoils the simple $\{\cos(\phi), \sin(\phi)\}$ dependence of the form factors.
Yet, as visible from the LCSR expressions, the numerically dominating effects are coming from the $\eT$ and $\eP$ decay constants entering the twist-2 contributions and essentially follow the $\{\cos(\phi), \sin(\phi)\}$ dependence of the FKS picture.
Essentially, the mixing angle can be considered as an effective mixing angle in the ratios of form factors, since various other effects associated with subleading effects (higher-twist DAs, 2-gluon contributions at NLO, \dots) induce a new dependence on $\eT-\eP$ mixing, not necessarily related to the single FKS angle $\phi$.
In particular, the U(1) anomaly \cref{eq:hqhs} entering at twist-3 and twist-4 level yields a large deviation from the simple picture.
Similarly, the inclusion of 2-gluon contributions is related to the non-trivial mixing which already shows up in the RGE for the decay constants at $O(\alpha_s)$.
However, both effects are numerically subleading, and the mixing behaviour is more or less governed by the simple FKS approximation.

For internal checks and the illustration of the $\phi$-dependence of the form factors, we evaluated numerically the ratios of \cref{eq:ctan} for all of the decays. 
We confirm that, within our framework, ratios of semileptonic $B_{(s)}, D_{(s)} \to \eT, \eP$ decays are suitable for the extraction of the mixing parameters, in particular the $\eT-\eP$ mixing angle.
Nonetheless, the extraction of the $\eT -\eP$ mixing angle from these ratios, due to the complex structure of the mixing in the form factors, can only be used as an indication of the general behaviour, but not as a method for a precise extraction of the $\phi$ angle.\footnote{
The authors of \cite{Colangelo:2010wg} were aware of the problem and have proposed a method for phenomenological extraction of the gluonic contributions by examining the semileptonic $D_s$ decays in combination with the two-body non-leptonic $D_s$ decays to $\eT$ and $\eP$.}

\section{Summary}

We presented an updated analysis of $B_{(s)}, D_{(s)} \rightarrow \eta^{(\prime)}$ form factors.
This update is motivated by recent experimental progress on the measurement of the semileptonic decays $B_{(s)}, D_{(s)} \rightarrow \eta^{(\prime)} \ell \nu$.
We use updated light-cone sum rule (LCSR) results and state-of-the-art $q^2$ parametrisations to determine the relevant transition form factors across the full kinematic range.
Our predictions agree well with the preliminary lattice estimate and with existing experimental extractions.
We also provide updated theory predictions for the semileptonic branching ratios, lepton-flavour universality ratios and forward-backwards asymmetries, and found a perfect agreement with the available experimental results.

Using all available experimental data, we extract the CKM matrix elements $|V_{ub}|$, $V_{cs}$, and $V_{cd}$.
We find that the resulting values are now competitive in precision with those from more traditional semileptonic decays.

We finally discuss the $\eta - \eta'$ mixing angle, which dominates the uncertainties of most of our predictions.
Although this angle can currently be extracted from data, it relies on a simplified mixing scheme, which will not be usable for precision in these decays.

This work brings the theoretical treatment of $B_{(s)}, D_{(s)} \to \eta^{(\prime)}$ transitions to a level comparable with that of better-studied modes like $B \to \pi$ and $D \to K$, therefore contributing to a more precise and robust flavour physics program.
Beyond their direct phenomenological value, these decays play a role in background modelling for other semileptonic processes, including $B \to \pi \ell \nu$ and inclusive $B \to X_u \ell \nu$ decays, and provide inputs relevant for understanding non-factorizable effects in hadronic $B$ and $D$ decays.

\section*{Acknowledgements}
We thank Domagoj Leljak and Danny van Dyk for their contribution at the early stages of the work.
Useful discussions with Thorsten Feldmann are greatly appreciated.
BM would like to acknowledge the support of the Croatian Science Foundation (HRZZ) under the project ``Nonperturbative QCD in heavy flavour physics'' (IP-2024-05-4427). 

\appendix

\section{Parameters used} \label{app:fixed_params}
In this section, we list the theoretical and experimental inputs used in our analysis.
For the quark masses, we use
\begin{gather*}
    \bar m_b(\bar m_b) = 4.18 \pm 0.03 \,\mathrm{GeV} \,,\\
    \bar m_c(\bar m_c) = 1.275 \pm 0.025 \,\mathrm{GeV} \,, \\
    \bar m_s(2\,\mathrm{GeV}) = 95 \pm 5 \,\mathrm{MeV} \,, \\
    \bar m_u(2\,\mathrm{GeV}) = 4.8^{+0.5}_{-0.3} \,\mathrm{MeV} \,, \\
    \bar m_d(2\,\mathrm{GeV}) = 2.3^{+0.7}_{-0.5} \,\mathrm{MeV} \,,
\end{gather*}
The meson decay constants are obtained from lattice estimates. We use
\begin{gather*}
    f_D = 212.0(0.7)\,\hbox{MeV~\cite{Carrasco:2014poa,Bazavov:2017lyh}} \,,\\
    f_{D_s} = 249.9(0.5)\,\hbox{MeV~\cite{Carrasco:2014poa,Bazavov:2017lyh}} \,,\\
    f_B = 190.0(1.3)\,\hbox{MeV~\cite{Bazavov:2017lyh,ETM:2016nbo,Dowdall:2013tga,Hughes:2017spc}} \,,\\
    f_{B_s} = 230.3(1.3)\,\hbox{MeV~\cite{Bazavov:2017lyh,ETM:2016nbo,Dowdall:2013tga,Hughes:2017spc}} \,.
\end{gather*}
The uncertainty associated with the above values is too small to have a noticeable impact on our results, and we fix these values in our numerical analysis.

For the CKM matrix elements $|V_{ub}|,V_{cs}$ and $V_{cd}$, we assumed the Wolfenstein parametrisation for the CKM matrix, using~\cite{Charles:2004jd,UTfit:2022hsi}
\begin{align}\label{eq:ckm}
    A          &= 0.81975 \pm 0.00645 \ ,  &\lambda&    = 0.22499 \pm 0.00022 \ , \\
    \bar{\rho} &= 0.1598  \pm 0.0076 \ ,   &\bar{\eta}& = 0.3548 \pm 0.0054 \ .
\end{align}
Meson masses correspond to EOS default values, which match the PDG ones~\cite{ParticleDataGroup:2024cfk}.
For the $\eta, \eta^{\prime}$ parameters, we used the lattice results of Ref.~\cite{Bali:2021qem}.
Resonances used in the BSZ parametrisation of the form factors are given in \cref{tab:resonances}.
The other inputs are listed in Ref.~\cite{Duplancic:2015zna}.

\begin{table}[ht]
    \centering
    \begin{tabular}{l cccc}
    \toprule
        $f$        & $b \to s$ & $b \to q$ & $c \to s$ & $c \to q$ \\
    \midrule
        $f^{+, T}$ & 5.415 GeV & 5.325 GeV & 2.0103 GeV & 2.1121 GeV \\
        $f^0$      & 5.63 GeV & 5.54 GeV & 2.318 GeV & 2.105 GeV \\
    \bottomrule
    \end{tabular}
    \caption{Masses of the resonances, $m_R$, used in the BSZ description of the form factors taken from Refs.~\cite{ParticleDataGroup:2024cfk,Lang:2015hza}.
    These parameters are considered fixed in our parameterisation.}
    \label{tab:resonances}
\end{table}

\section{Experimental results used in the fits} \label{app:exp_inputs}
This section summarizes the available experimental results for charged $B, D_{(s)}\to\eta^{(\prime)}\ell\bar\nu_\ell$ decays.
For neutral decays, only 90\% CL limits have been set
\begin{eqnarray}
    \mathcal{B}(D^0\to\eta ee) & < 3 \times 10^{-6}\hbox{~\cite{BESIII:2018hqu,CLEO:1996jxx}}, \\
    \mathcal{B}(D^0\to\eta\mu\mu) & < 5.3 \times 10^{-4}\hbox{~\cite{CLEO:1996jxx}}, \\
    \mathcal{B}(B^0\to\eta ee) & < 1.05 \times 10^{-7}\hbox{~\cite{Belle:2024cis}}, \\
    \mathcal{B}(B^0\to\eta\mu\mu) & < 9.4 \times 10^{-8}\hbox{~\cite{Belle:2024cis}}.
\end{eqnarray}
Due to the complexity of neutral decays, we don't use these limits in our analysis.

All the charged modes have been seen, and the branching ratios are usually known from several experiments.
When available, we use the combinations of the PDG~\cite{ParticleDataGroup:2024cfk}, which usually match those of HFLAV~\cite{HeavyFlavorAveragingGroupHFLAV:2024ctg}.
We also updated some of these averages with the newest measurements using the PDG averaging procedure.
For some of these decays, information on the differential branching ratio is also available via correlated binned measurements.
To avoid any D'Agostini bias~\cite{DAgostini:1993arp}, we convert these binned measurements to kinematic PDFs by normalising to the sum of the bins and by dropping the last bin.
All experimental constraints are available in EOS and can be summarised as follows.

\begin{description}
    \item[$\boldsymbol{D\to\eta\ell\nu}$] For the total branching ratios we used $\mathcal{B}(D\to\eta e\nu) = (9.99 \pm 0.58) \times 10^{-4} \ [S=1.55]$~\cite{CLEO:2010pjh,BESIII:2025hjc} and $\mathcal{B}(D\to\eta\mu\nu) = (9.08 \pm 0.35 \pm 0.23) \times 10^{-4}$~\cite{BESIII:2025hjc}.
    The differential branching ratios are also available in four $q^2$ bins for the $e$ and $\mu$ modes~\cite{BESIII:2025hjc}, although not enough information is provided to fully reconstruct the covariance matrix. We therefore neglected the correlations between these two modes.
    \item[$\boldsymbol{D\to\eta'\ell\nu}$] For the total branching ratios we used $\mathcal{B}(D\to\eta' e\nu) = (1.84 \pm 0.18) \times 10^{-4}$~\cite{CLEO:2010pjh,Ablikim:2020hsc,BESIII:2024njj} and $\mathcal{B}(D\to\eta'\mu\nu) = (1.92 \pm 0.28 \pm 0.08) \times 10^{-4}$~\cite{ParticleDataGroup:2024cfk}.
    The differential branching ratios are also available in four $q^2$ bins for the $e$ and $\mu$ modes~\cite{BESIII:2024njj}.
    \item[$\boldsymbol{D_s\to\eta\ell\nu}$] For the total branching ratios we used $\mathcal{B}(D_s\to\eta e\nu) = (2.26 \pm 0.06) \times 10^{-2}$~\cite{ParticleDataGroup:2024cfk} and $\mathcal{B}(D_s\to\eta\mu\nu) = (2.24 \pm 0.07) \times 10^{-2}$~\cite{BESIII:2017ikf,BESIII:2023gbn}.
    The differential branching ratios are also available in eight $q^2$ bins for the $e$~\cite{BESIII:2023ajr} and $\mu$ modes~\cite{BESIII:2023gbn}, where we combined the $\eta$ decay modes and assumed $100\%$ correlated systematic uncertainties between the two analyses.
    \item[$\boldsymbol{D_s\to\eta'\ell\nu}$] For the total branching ratios we used $\mathcal{B}(D_s\to\eta' e\nu) = (8.0 \pm 0.4) \times 10^{-3}$~\cite{ParticleDataGroup:2024cfk} and $\mathcal{B}(D_s\to\eta'\mu\nu) = (8.0 \pm 0.6) \times 10^{-3}$~\cite{BESIII:2017ikf,BESIII:2023gbn}.
    The differential branching ratios are also available in three $q^2$ bins for the $e$~\cite{BESIII:2023ajr} and $\mu$ modes~\cite{BESIII:2023gbn}, and we assumed $100\%$ correlated systematic uncertainties between the two analyses.
    \item[$\boldsymbol{B\to\eta\ell\nu}$] For the total branching ratios we used $\mathcal{B}(B\to\eta\ell\nu) = (3.44 \pm 0.43) \times 10^{-5}$~\cite{HeavyFlavorAveragingGroupHFLAV:2024ctg} which averages the two light lepton modes.
    The differential branching ratios are also available in five $q^2$ bins, averaged over the two lepton modes~\cite{BaBar:2012thb}.
    \item[$\boldsymbol{B\to\eta'\ell\nu}$] For the total branching ratios we used $\mathcal{B}(B\to\eta'\ell\nu) = (2.49 \pm 0.67) \times 10^{-5}$~\cite{HeavyFlavorAveragingGroupHFLAV:2024ctg} which averages the two light lepton modes.
\end{description}

\section{Experimental distributions}\label{app:distribution_results}
\begin{figure}[ht]
    \centering
    \includegraphics[width=0.49\linewidth]{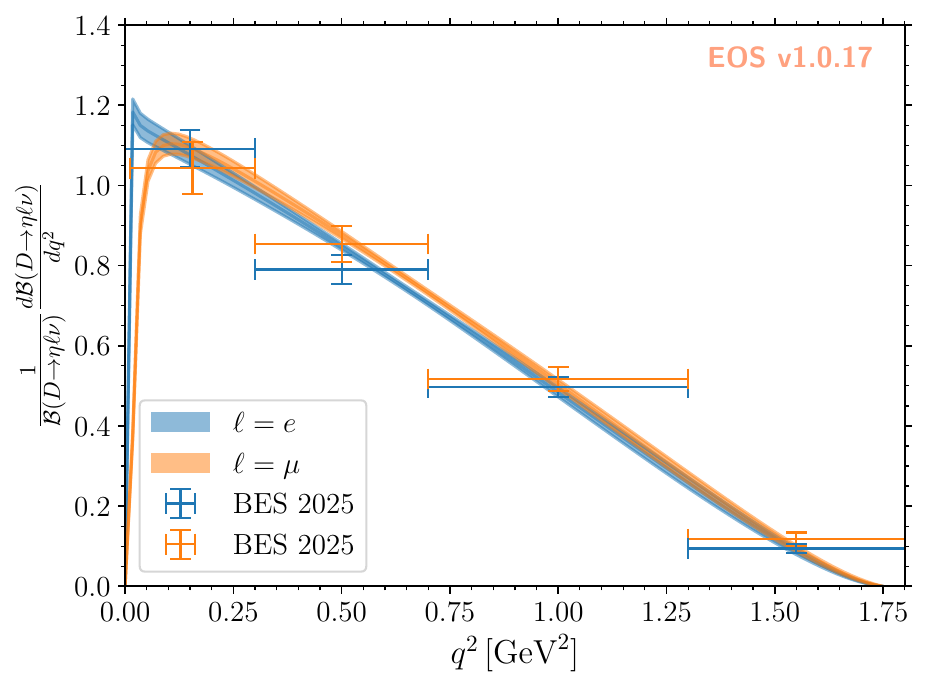}
    \includegraphics[width=0.49\linewidth]{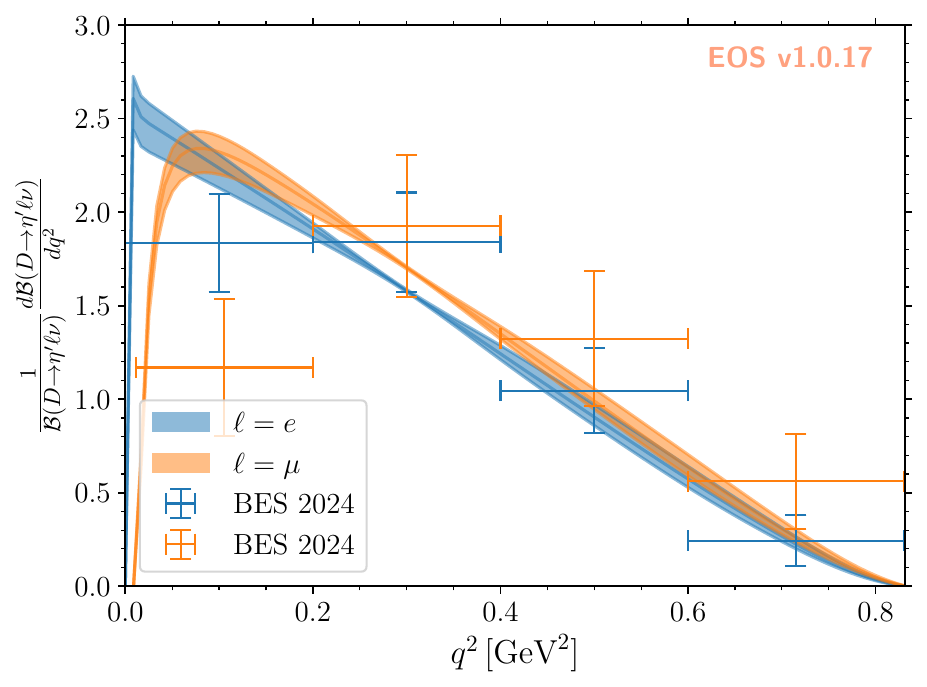} \\
    \includegraphics[width=0.49\linewidth]{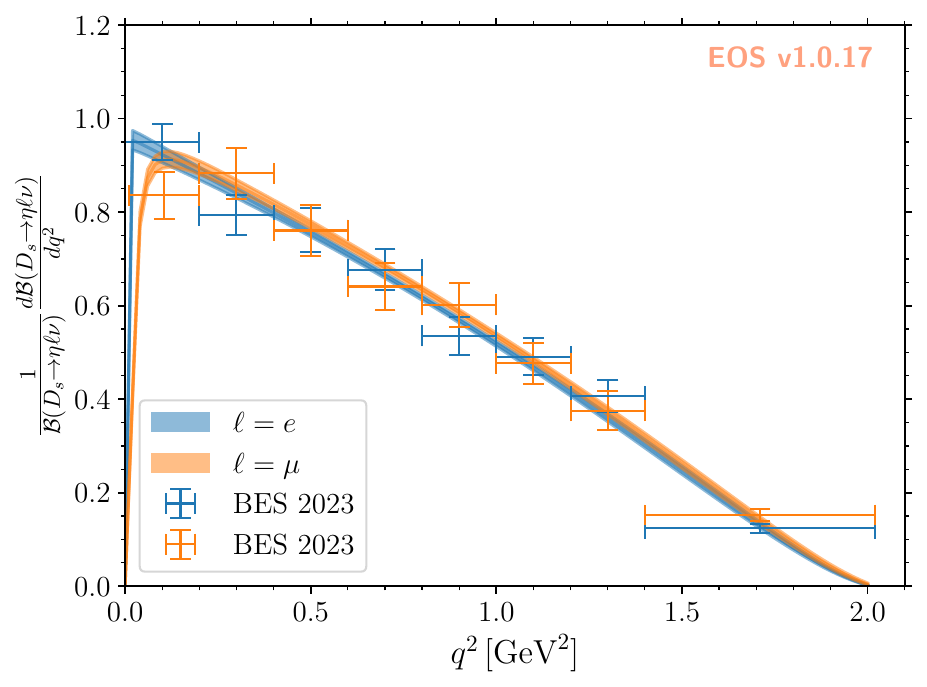}
    \includegraphics[width=0.49\linewidth]{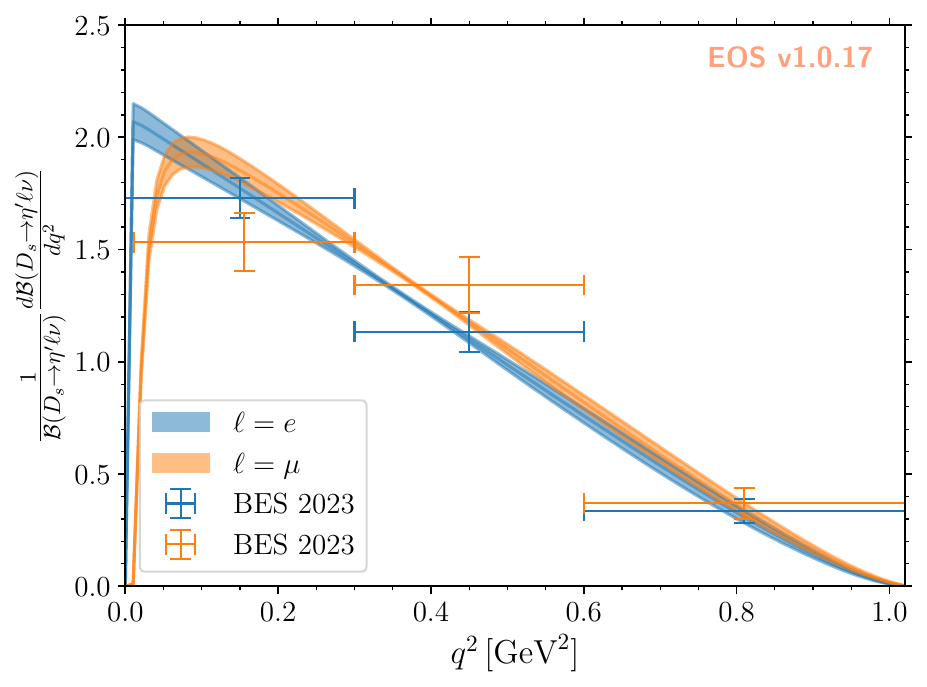} \\
    \includegraphics[width=0.49\linewidth]{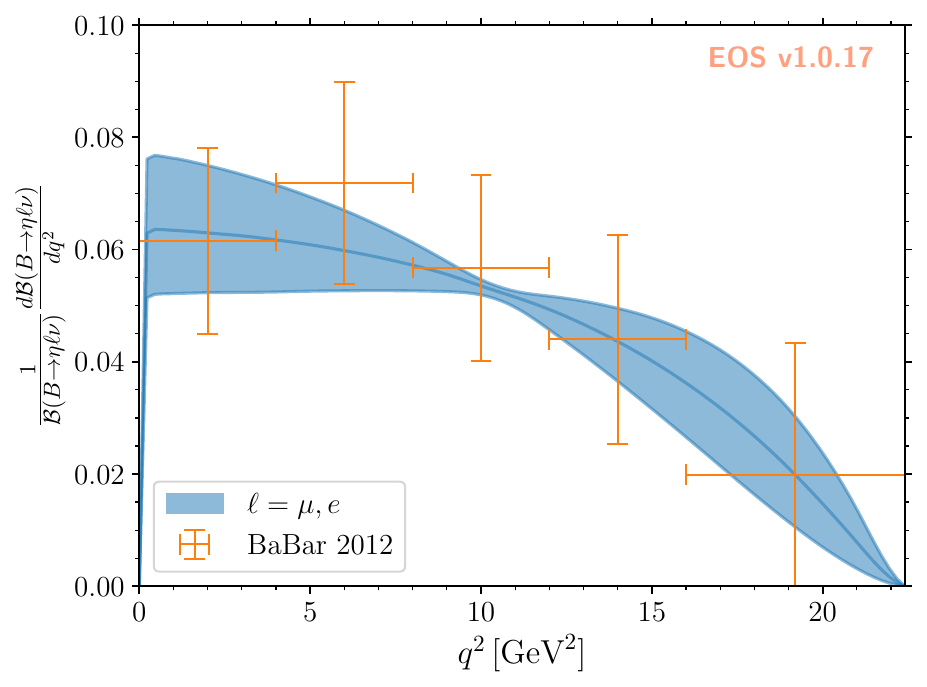}
    \caption{Comparison between our fit results for differential BRs and some experimental measurements (listed in \cref{app:exp_inputs}).
    The bands correspond to $1\sigma$ uncertainty intervals for the normalised differential branching ratios.}
    \label{fig:experimental_distributions}
\end{figure}

In \cref{fig:experimental_distributions}, we provide comparison plots between our predicted differential branching ratios and the available experimental results.
As expected from the excellent $p$-values of the fits, we find no tension between the data and our results.
The peculiar shape of the uncertainty band, particularly visible in the $B\to\eta\ell\nu$ plot, is due to the balance between the large increase of our form factor uncertainties at high $q^2$ and the closure of the phase space.

\section{Coefficients of the form-factor expansion}\label{app:z_coeff}
In this section, we provide the coefficients of the form factor expansion \cref{eq:SSE}.
The posterior distributions are perfectly described by multivariate normal distributions, for which we provide means and covariance.

\begin{table}[ht]
\resizebox{\textwidth}{!}{
    \centering
    \begin{tabular}{cccccccc}
        \toprule
        $\alpha_0^+$ & $\alpha_1^+$ & $\alpha_2^+$ & $\alpha_1^0$ & $\alpha_2^0$ & $\alpha_0^T$ & $\alpha_1^T$ & $\alpha_2^T$ \\
        \midrule
         $0.19 \pm 0.05$ & $-0.50 \pm 0.23$ & $0.44 \pm 2.89$ & $0.06 \pm 0.15$ & $0.18 \pm 3.02$ & $0.19 \pm 0.05$ & $-0.54 \pm 0.25$ & $0.69 \pm 3.24$ \\
        \midrule
         $1.0000$ & $-0.7376$ & $-0.0483$ & $-0.1548$ & $-0.1003$ & $0.9888$ & $-0.7386$ & $0.0422$ \\
                 & $1.0000$ & $-0.4074$ & $0.3377$ & $-0.0051$ & $-0.7415$ & $0.7018$ & $-0.0695$ \\
                 &         & $1.0000$ & $-0.1669$ & $0.4366$ & $0.0101$ & $-0.0443$ & $0.0396$ \\
                 &         &         & $1.0000$ & $-0.4540$ & $-0.1700$ & $0.3528$ & $-0.0577$ \\
                 &         &         &         & $1.0000$ & $-0.0351$ & $0.0424$ & $0.0179$ \\
                 &         &         &         &         & $1.0000$ & $-0.7263$ & $-0.0325$ \\
                 &         &         &         &         &         & $1.0000$ & $-0.4214$ \\
                 &         &         &         &         &         &         & $1.0000$ \\
        \bottomrule
    \end{tabular}}
    \caption{Multinormal approximation of the posterior distribution for the $B\to\eta$ form factor parameters.}
\end{table}

\begin{table}[ht]
\resizebox{\textwidth}{!}{
    \centering
    \begin{tabular}{cccccccc}
        \toprule
        $\alpha_0^+$ & $\alpha_1^+$ & $\alpha_2^+$ & $\alpha_1^0$ & $\alpha_2^0$ & $\alpha_0^T$ & $\alpha_1^T$ & $\alpha_2^T$ \\
        \midrule
         $0.14 \pm 0.05$ & $-0.43 \pm 0.21$ & $0.55 \pm 2.88$ & $0.08 \pm 0.14$ & $0.20 \pm 2.99$ & $0.16 \pm 0.05$ & $-0.49 \pm 0.25$ & $0.78 \pm 3.43$ \\
        \midrule
         $1.0000$ & $-0.7632$ & $0.0032$ & $0.1051$ & $-0.0726$ & $0.9859$ & $-0.6815$ & $0.0635$ \\
                 & $1.0000$ & $-0.4271$ & $0.1589$ & $-0.0261$ & $-0.7762$ & $0.6958$ & $-0.0784$ \\
                 &         & $1.0000$ & $-0.1478$ & $0.4260$ & $0.0513$ & $-0.0769$ & $0.0432$ \\
                 &         &         & $1.0000$ & $-0.4333$ & $0.0466$ & $0.2651$ & $-0.0517$ \\
                 &         &         &         & $1.0000$ & $-0.0249$ & $0.0311$ & $0.0311$ \\
                 &         &         &         &         & $1.0000$ & $-0.7021$ & $0.0024$ \\
                 &         &         &         &         &         & $1.0000$ & $-0.4394$ \\
                 &         &         &         &         &         &         & $1.0000$ \\
        \bottomrule
    \end{tabular}}
    \caption{Multinormal approximation of the posterior distribution for the $B\to\eta^\prime$ form factor parameters.}
\end{table}

\begin{table}[ht]
\resizebox{\textwidth}{!}{
    \centering
    \begin{tabular}{cccccccc}
        \toprule
        $\alpha_0^+$ & $\alpha_1^+$ & $\alpha_2^+$ & $\alpha_1^0$ & $\alpha_2^0$ & $\alpha_0^T$ & $\alpha_1^T$ & $\alpha_2^T$ \\
        \midrule
         $-0.25 \pm 0.01$ & $0.61 \pm 0.20$ & $-0.48 \pm 4.08$ & $-0.11 \pm 0.20$ & $-0.50 \pm 4.16$ & $-0.25 \pm 0.01$ & $0.65 \pm 0.22$ & $-0.56 \pm 4.52$ \\
        \midrule
         $1.0000$ & $0.1552$ & $-0.3990$ & $0.2888$ & $-0.3831$ & $0.6856$ & $0.1344$ & $-0.0702$ \\
                 & $1.0000$ & $-0.5423$ & $0.3240$ & $-0.0096$ & $0.0883$ & $0.3259$ & $0.0043$ \\
                 &         & $1.0000$ & $-0.1182$ & $0.4392$ & $-0.0770$ & $0.0081$ & $0.0082$ \\
                 &         &         & $1.0000$ & $-0.4108$ & $0.1802$ & $0.3413$ & $-0.0042$ \\
                 &         &         &         & $1.0000$ & $-0.0487$ & $0.0812$ & $0.0122$ \\
                 &         &         &         &         & $1.0000$ & $0.2107$ & $-0.4797$ \\
                 &         &         &         &         &         & $1.0000$ & $-0.5239$ \\
                 &         &         &         &         &         &         & $1.0000$ \\
        \bottomrule
    \end{tabular}}
    \caption{Multinormal approximation of the posterior distribution for the $B_s\to\eta$ form factor parameters.}
\end{table}

\begin{table}[ht]
\resizebox{\textwidth}{!}{
    \centering
     \begin{tabular}{cccccccc}
        \toprule
        $\alpha_0^+$ & $\alpha_1^+$ & $\alpha_2^+$ & $\alpha_1^0$ & $\alpha_2^0$ & $\alpha_0^T$ & $\alpha_1^T$ & $\alpha_2^T$ \\
        \midrule
         $0.28 \pm 0.02$ & $-0.79 \pm 0.27$ & $1.06 \pm 5.88$ & $0.14 \pm 0.27$ & $0.58 \pm 5.94$ & $0.31 \pm 0.02$ & $-0.92 \pm 0.31$ & $1.18 \pm 7.00$ \\
        \midrule
         $1.0000$ & $-0.0111$ & $-0.2404$ & $0.3662$ & $-0.2279$ & $0.8457$ & $0.0784$ & $-0.0607$ \\
                 & $1.0000$ & $-0.5833$ & $0.2816$ & $-0.0392$ & $-0.0132$ & $0.3135$ & $-0.0135$ \\
                 &         & $1.0000$ & $-0.1258$ & $0.4303$ & $-0.0730$ & $-0.0236$ & $0.0304$ \\
                 &         &         & $1.0000$ & $-0.4241$ & $0.2993$ & $0.3227$ & $-0.0267$ \\
                 &         &         &         & $1.0000$ & $-0.0372$ & $0.0664$ & $0.0079$ \\
                 &         &         &         &         & $1.0000$ & $0.1619$ & $-0.3571$ \\
                 &         &         &         &         &         & $1.0000$ & $-0.5665$ \\
                 &         &         &         &         &         &         & $1.0000$ \\
        \bottomrule
    \end{tabular}}
    \caption{Multinormal approximation of the posterior distribution for the $B_s\to\eta^\prime$ form factor parameters.}
\end{table}

\begin{table}[ht]
\resizebox{\textwidth}{!}{
    \centering
    \begin{tabular}{cccccccc}
        \toprule
        $\alpha_0^+$ & $\alpha_1^+$ & $\alpha_2^+$ & $\alpha_1^0$ & $\alpha_2^0$ & $\alpha_0^T$ & $\alpha_1^T$ & $\alpha_2^T$ \\
        \midrule
         $0.38 \pm 0.13$ & $-0.51 \pm 0.48$ & $-1.21 \pm 17.52$ & $0.18 \pm 0.49$ & $-0.24 \pm 16.54$ & $0.38 \pm 0.13$ & $-0.67 \pm 0.53$ & $-1.98 \pm 19.34$ \\
        \midrule
         $1.0000$ & $-0.2108$ & $-0.1158$ & $0.2648$ & $-0.0943$ & $0.9917$ & $-0.2848$ & $-0.0783$ \\
                 & $1.0000$ & $-0.6087$ & $0.0520$ & $-0.1392$ & $-0.2205$ & $0.1622$ & $-0.0174$ \\
                 &         & $1.0000$ & $-0.1076$ & $0.4443$ & $-0.0828$ & $-0.1166$ & $0.1586$ \\
                 &         &         & $1.0000$ & $-0.5896$ & $0.2595$ & $-0.0203$ & $0.0040$ \\
                 &         &         &         & $1.0000$ & $-0.0564$ & $-0.0598$ & $0.0826$ \\
                 &         &         &         &         & $1.0000$ & $-0.2499$ & $-0.1486$ \\
                 &         &         &         &         &         & $1.0000$ & $-0.6582$ \\
                 &         &         &         &         &         &         & $1.0000$ \\
        \bottomrule
    \end{tabular}}
    \caption{Multinormal approximation of the posterior distribution for the $D\to\eta$ form factor parameters.}
\end{table}

\begin{table}[ht]
\resizebox{\textwidth}{!}{
    \centering
    \begin{tabular}{cccccccc}
        \toprule
        $\alpha_0^+$ & $\alpha_1^+$ & $\alpha_2^+$ & $\alpha_1^0$ & $\alpha_2^0$ & $\alpha_0^T$ & $\alpha_1^T$ & $\alpha_2^T$ \\
        \midrule
         $0.28 \pm 0.12$ & $-0.62 \pm 0.62$ & $-2.89 \pm 22.58$ & $0.97 \pm 0.55$ & $-9.65 \pm 23.89$ & $0.34 \pm 0.14$ & $-0.87 \pm 0.85$ & $-6.45 \pm 29.98$ \\
        \midrule
         $1.0000$ & $-0.5750$ & $-0.1383$ & $0.5938$ & $-0.2741$ & $0.9935$ & $-0.6005$ & $-0.1287$ \\
                 & $1.0000$ & $-0.4705$ & $-0.2608$ & $0.0373$ & $-0.5969$ & $0.5111$ & $0.0444$ \\
                 &         & $1.0000$ & $-0.1939$ & $0.4416$ & $-0.1076$ & $0.0177$ & $0.0920$ \\
                 &         &         & $1.0000$ & $-0.6221$ & $0.5720$ & $-0.2861$ & $-0.0953$ \\
                 &         &         &         & $1.0000$ & $-0.2442$ & $0.1367$ & $0.0884$ \\
                 &         &         &         &         & $1.0000$ & $-0.6019$ & $-0.1842$ \\
                 &         &         &         &         &         & $1.0000$ & $-0.4237$ \\
                 &         &         &         &         &         &         & $1.0000$ \\
        \bottomrule
    \end{tabular}}
    \caption{Multinormal approximation of the posterior distribution for the $D\to\eta^\prime$ form factor parameters.}
\end{table}

\begin{table}[ht]
\resizebox{\textwidth}{!}{
    \centering
    \begin{tabular}{cccccccc}
        \toprule
        $\alpha_0^+$ & $\alpha_1^+$ & $\alpha_2^+$ & $\alpha_1^0$ & $\alpha_2^0$ & $\alpha_0^T$ & $\alpha_1^T$ & $\alpha_2^T$ \\
        \midrule
         $-0.47 \pm 0.02$ & $0.53 \pm 0.61$ & $4.17 \pm 23.08$ & $0.45 \pm 0.64$ & $0.24 \pm 23.66$ & $-0.45 \pm 0.02$ & $0.38 \pm 0.69$ & $10.32 \pm 24.63$ \\
        \midrule
         $1.0000$ & $0.1445$ & $-0.4629$ & $0.2070$ & $-0.4869$ & $0.4742$ & $0.0109$ & $-0.0442$ \\
                 & $1.0000$ & $-0.6531$ & $0.1576$ & $-0.1324$ & $-0.0121$ & $0.1828$ & $-0.0368$ \\
                 &         & $1.0000$ & $-0.1249$ & $0.3903$ & $-0.0068$ & $-0.0031$ & $0.0047$ \\
                 &         &         & $1.0000$ & $-0.6579$ & $0.0635$ & $0.1846$ & $-0.0326$ \\
                 &         &         &         & $1.0000$ & $-0.0341$ & $-0.0076$ & $-0.0036$ \\
                 &         &         &         &         & $1.0000$ & $0.1035$ & $-0.4892$ \\
                 &         &         &         &         &         & $1.0000$ & $-0.6279$ \\
                 &         &         &         &         &         &         & $1.0000$ \\
        \bottomrule
    \end{tabular}}
    \caption{Multinormal approximation of the posterior distribution for the $D_s\to\eta$ form factor parameters.}
\end{table}

\begin{table}[ht]
\resizebox{\textwidth}{!}{
    \centering
    \begin{tabular}{cccccccc}
        \toprule
        $\alpha_0^+$ & $\alpha_1^+$ & $\alpha_2^+$ & $\alpha_1^0$ & $\alpha_2^0$ & $\alpha_0^T$ & $\alpha_1^T$ & $\alpha_2^T$ \\
        \midrule
         $0.50 \pm 0.05$ & $-0.72 \pm 0.92$ & $-10.99 \pm 44.85$ & $0.43 \pm 1.02$ & $-11.03 \pm 49.49$ & $0.58 \pm 0.06$ & $-0.17 \pm 1.29$ & $-28.07 \pm 58.38$ \\
        \midrule
         $1.0000$ & $-0.0884$ & $-0.1844$ & $0.4339$ & $-0.4366$ & $0.9033$ & $0.1276$ & $-0.1005$ \\
                 & $1.0000$ & $-0.6888$ & $0.0768$ & $-0.1259$ & $-0.1676$ & $0.1432$ & $-0.0273$ \\
                 &         & $1.0000$ & $-0.1679$ & $0.3924$ & $-0.0274$ & $-0.0598$ & $0.0233$ \\
                 &         &         & $1.0000$ & $-0.7341$ & $0.3434$ & $0.2689$ & $-0.1212$ \\
                 &         &         &         & $1.0000$ & $-0.2669$ & $-0.1572$ & $0.0793$ \\
                 &         &         &         &         & $1.0000$ & $0.1237$ & $-0.2951$ \\
                 &         &         &         &         &         & $1.0000$ & $-0.6796$ \\
                 &         &         &         &         &         &         & $1.0000$ \\
        \bottomrule
    \end{tabular}}
    \caption{Multinormal approximation of the posterior distribution for the $D_s\to\eta^\prime$ form factor parameters.}
\end{table}

\bibliography{References_eta-etaP}
\bibliographystyle{JHEP}

\end{document}